\documentclass[lettersize,journal]{IEEEtran}

\usepackage[T1]{fontenc}
\usepackage{siunitx}
\usepackage{cite}
\usepackage[cmex10]{amsmath}
\usepackage{amssymb}
\usepackage{makecell}
\usepackage{boldline}
\usepackage{adjustbox}
\usepackage{caption}
\usepackage{tikz}
\usetikzlibrary{shapes,arrows,fit,positioning,shadows,calc}
\usetikzlibrary{plotmarks}
\usetikzlibrary{decorations.pathreplacing}
\usetikzlibrary{patterns}
\usepackage{algorithm}
\usepackage{algpseudocode}
\algnewcommand{\Initialization}[1]{%
  \State \textbf{initialization:}
  \Statex \hspace*{\algorithmicindent}\parbox[t]{.8\linewidth}{\raggedright #1}
}
\usetikzlibrary{automata}
\usepackage{xcolor}
\usetikzlibrary{spy}

\usepackage{pgfplots}
\pgfplotsset{compat=newest}

\usepackage{graphicx}
\usepackage{ragged2e}
\usepackage{hhline}
\usepackage{xfrac}

\newcommand{\pr}[1]{\ensuremath{\left[#1\right]}}
\newcommand{\pc}[1]{\ensuremath{\left(#1\right)}}
\newcommand{\chav}[1]{\ensuremath{\left\{#1\right\}}}
\newcommand{\PM}[1]{\ensuremath{\left|#1\right|}}

\definecolor{b}{rgb}{0, 0, 1}
\definecolor{r0}{rgb}{0, 0, 0}
\definecolor{r}{rgb}{1, 0, 0}
\definecolor{r1}{rgb}{0,0,0}
\definecolor{r2}{rgb}{0,0,0}
\definecolor{r3}{rgb}{0,0,0}
\definecolor{r4}{rgb}{1,0,0}

\definecolor{dark_green}{rgb}{0, 0.33, 0.13}
\definecolor{naplesyellow}{rgb}{0.99, 0.93, 0.0}
\definecolor{aureolin}{rgb}{1, 0.8, 0}
\definecolor{purple}{rgb}{0.4940 0.1840 0.5560}

\ifCLASSINFOpdf
\else
\fi
\usepackage{amsmath}
\usepackage[cmintegrals]{newtxmath}
\begin{document}
%

%
\title{Power Minimization under Quality of Service Constraints for  MIMO Systems with a RIS-based Transmitter}

%

\author{Erico~S.~P.~Lopes,~\IEEEmembership{Graduate Student Member,~IEEE},
				~{Lukas~T.~N.~Landau,~\IEEEmembership{Senior Member,~IEEE},~and~{Amine~Mezghani,~\IEEEmembership{Member,~IEEE}}}
\thanks{E.~S.~P.~Lopes is with the National Institute of Industrial Property of Brazil (INPI) (e-mail: lopespserico@gmail.com) and with the Department of Electrical Engineering, PUC-Rio, Rio de Janeiro. L.~T.~N.~Landau is with CETUC and the Department of Electrical Engineering, PUC-Rio, Rio de Janeiro, 
Brazil, (e-mail: landau@puc-rio.br).
A.~Mezghani is with the Department of Electrical and Computer Engineering, University of Manitoba, Winnipeg, MB,
R3T 5V6, Canada, (e-mail: amine.mezghani@umanitoba.ca).
This work was supported by FAPERJ, FAPESP, CNPq, and, CAPES. Parts of the work were presented in \textit{2023 IEEE 9th International Workshop on Computational Advances in Multi-Sensor Adaptive Processing (CAMSAP)} \cite{lopes_camsap2023}; in \textit{2024 {IEEE} Int. Conf. Commun. ({ICC})} \cite{lopes_icc2024}, and in \textit{Int. Symp. on Wireless Comm. Systems (ISWCS) 2024} \cite{lopes_iswcs2024}.}
}

\maketitle

\vspace{-5em}

\begin{abstract}
\textcolor{r1}{This study investigates a virtual multiuser multiple-input multiple-output (MU-MIMO) system with PSK modulation, realized with a reconfigurable intelligent surface (RIS)-based transmitter. The study focuses on minimizing transmit power under quality-of-service (QoS) constraints while addressing the associated computational complexity. A discrete phase-shift RIS model is considered, and the power minimization problem is formulated in two scenarios. First, for QPSK user data, the symbol-error probability (SEP) is adopted as the QoS criterion. Second, for general $M$-PSK modulation, the union-bound SEP (UBSEP) is used to define the QoS constraints.
Based on the considered formulations, a partial branch-and-bound (PBB) approach is developed, which improves on full branch-and-bound (FBB) methods in the sense of allowing for favorable complexity performance trade-offs. For the special case of high-resolution RIS, the discrete phase-shift set is approximated by its continuous counterpart, enabling the reformulation of the original problems as constrained optimizations on an oblique manifold, which are solved with reduced computational complexity with the proposed bisection method. 
Numerical results demonstrate the effectiveness of the proposed approaches in minimizing the transmit power for different SEP requirements and showcase the balance between power efficiency and computational complexity.}
\end{abstract}



%
\IEEEpeerreviewmaketitle

\section{Introduction}

To enable the foreseen applications of future generations of wireless communications, achieving strict \textcolor{r3}{quality of service (QoS)} requirements, e.g., low latency, high reliability, and high data rate \cite{Ris_6G,rappaport_acess2019}, are necessary. According to \cite{6G_Future_Directions}, the next generation is expected to require a 100-fold improvement in data rates for the uplink and a 50-fold improvement for the downlink while achieving 10 thousand times higher reliability when compared to 5G. \textcolor{r3}{Multiuser multiple-input multiple-output (MU-MIMO)} systems are a key technology for meeting these requirements \cite{6G_Future_Directions}. Yet, equipping base stations (BSs) with large antenna arrays increases hardware costs and energy consumption, creating bottlenecks for practical implementation.
In this context, \textcolor{r3}{reconfigurable intelligent surfaces (RISs)} have emerged as a promising technology for beyond-5G/6G wireless communications as it can improve the energy efficiency of wireless systems while offering the benefits of low-cost and easy integration into the currently deployed wireless systems \cite{rappaport_acess2019,Qingqing_ComMag2020}. 
In essence, a RIS comprises an array of reconfigurable reflective
elements where the reflection coefficient of each element is real-time, electrically controllable. By adjusting the RIS elements' reflection coefficients, one can perform passive signal shaping without necessitating a power amplifier, which yields an advantage in terms of energy efficiency when compared with conventional MIMO BSs \cite{Liu_tutorial2021,Ngo_tcom2013}. With this, RIS has become popular in the literature of mmWave and multi-antenna communications, most commonly for backscatter communications \cite{zhao2020metasurface,galappaththige2022ris}, wireless propagation environment control \cite{DiRenzo_twc2021,Alexandropoulos_tvt2024}, and beamforming \cite{jung2021optimality,lin2020reconfigurable}. Regarding beamforming for RIS, recent advances showed that RIS is also able to perform simultaneous passive beamforming and transmit physical information \cite{Wenjing_jsac2020,Lin_gcom2020}, which led to the development of different studies for this context. 
The work from  \cite{Qiang_Wcom2021} compares RIS-based modulation with spatial multiplexing and discusses the former's benefits over the latter. In \cite{Amine_Mis_tcom2022} the authors introduce the concept of modulating intelligent surfaces as RIS capable of performing passive beamforming for users served by a BS, embedding information through backscatter communication, or doing both simultaneously. The work from \cite{Karasik_tcom2020} demonstrates the advantage of using RIS for modulation over traditional beamforming, where RIS phase shifts are independent of transmitted information. \textcolor{r1}{The works in \cite{Basar_access2019,renzo2019smart} discuss how RIS enables signal modulation by dynamically controlling the reflection coefficients of its elements. For the specific case of PSK modulation, the RIS adjusts the phase of these reflection coefficients in real-time, based on the data vector. This introduces data-dependent phase shifts into the reflected signals, effectively realizing a PSK modulation on the transmitted signal.}

\textcolor{r1}{Similar to \cite{Karasik_tcom2020} and \cite{Basar_access2019}, this study utilizes the RIS modulation capabilities with the RIS-based transmitter setup, which realizes low-cost energy-efficient massive MIMO. As proposed in \cite{Karasik_tcom2020,Basar_access2019}, an efficient transmitter can be realized by illuminating an RIS with an unmodulated carrier signal generated by a nearby radio-frequency (RF) signal generator and changing the parameters of the reflecting elements to modulate and transmit information symbols. With this, the considered RIS-based transmitter realizes virtual MIMO systems with a single RF chain and cost-effective reflecting elements, which benefits the implementation of massive MIMO with reduced hardware complexity and increased energy efficiency. By employing the RIS-based transmission setup the optimization of the transmit signal can be done utilizing a similar mechanism as in symbol-level precoding (SLP) \cite{General_MMDDT_BB,lopes2021discrete,lopes_wcl2022,lopes_tcom2023,masouros_twc2018,Mingjie_framework,Mingjie_wcl2020}, which achieves high performance by varying the precoder for each symbol vector. Note that, although the SLP mechanism is also applied to the considered framework, the works from \cite{General_MMDDT_BB,lopes2021discrete,lopes_wcl2022,lopes_tcom2023,masouros_twc2018,Mingjie_framework,Mingjie_wcl2020} consider a BS equipped with one RF chain per antenna. In this sense, the considered setup can be favorable in terms of energy efficiency since only one RF chain is required. Moreover, once the algorithms considered in \cite{General_MMDDT_BB,lopes2021discrete,lopes_wcl2022,lopes_tcom2023,masouros_twc2018,Mingjie_framework,Mingjie_wcl2020} can be adapted to the considered setup, the utilization of RIS-based transmitters does not yield an increase in computational complexity, and similar QoS guarantees are attainable.}

Different works have arisen considering RIS-based transmission schemes. In \cite{li2022reconfigurable}, the authors jointly optimize the total power reflected from the RIS and the power allocation fraction assigned to each user. In \cite{Amine_Mis_tcom2022} to maximize each user’s spectral efficiency, the authors propose a joint non-convex optimization problem using the sum minimum mean-square error criterion. \textcolor{r1}{In \cite{Rongfang_2024}, the authors propose a power minimization problem for an uplink RIS-aided MIMO-NOMA system. In \cite{liu2021intelligent}, high-resolution RIS is considered, and the authors propose a power minimization problem under 
 minimum distance to the decision threshold (MDDT) constraints.}

Following the path of \cite{liu2021intelligent}, this study proposes a power minimization problem under QoS requisites. Yet, different than in \cite{liu2021intelligent}, the present study focuses on the minimization of the power radiated by the RF generator under the condition that the \textcolor{r3}{symbol-error probability (SEP)} of the users is below a given requisite. The discrete phase-shift RIS model is considered such that the reflecting elements' coefficients are restricted to a discrete set. The main contributions are delineated as follows.
\begin{itemize}
    \item For the case of QPSK users' data, where the SEP can be expressed with tabled functions \cite{lopes_tcom2023}, the study proposes the power minimization under SEP constraints (PSEP) problem. For the general case of $M$-PSK users' data, the utilization of the SEP would lead to constraint functions that require evaluation via Monte Carlo methods. With this, the PSEP problem is reformulated in the sense of substituting the SEP by the \textcolor{r3}{union-bound SEP (UBSEP)} in the constraint functions, resulting in the problem of power minimization under UBSEP constraints (PUBSEP).
    \item Based on the PSEP and PUBSEP problems, we develop a \textcolor{r3}{partial branch-and-bound (PBB)} algorithm that differs from conventional \textcolor{r3}{full branch-and-bound (FBB)} methods by accepting any solution that either attains the given target power budget of the system or is sufficiently close to the optimal such that it is considered unnecessary to continue the search process. 
    \item For the case where the number of discrete phase-shift RIS is large, i.e., for a high-resolution RIS, a reduced complexity method is proposed by approximating the discrete feasible set to its continuous counterpart. This leads to reformulations of the proposed PSEP and PUBSEP problems as constrained optimization problems on an oblique manifold. 
    \item The constrained optimization problems on an oblique manifold are solved via bisection methods (BM) that successively adjust the transmit power while evaluating the feasibility of the QoS constraints by solving, via the Riemannian conjugate gradient (RCG) algorithm, an auxiliary problem dependent only on the coefficients of the RIS reflecting elements.
    \item Finally, all proposed formulations are expanded for the case of imperfect channel state information (CSI) by considering that the worst-case CSI mismatch is limited in terms of its norm being smaller or equal to a given value of $\epsilon_k$ defined for each user. 
\end{itemize}
Numerical results underline that the proposed PBB method achieves significant complexity reduction with minor transmit power increase. Compared with \cite{liu2021intelligent}, the proposed techniques yield similar complexity with reduced transmit power.

\subsection{Remainder and Notation}

The remainder of this paper is organized as follows: Section~\ref{sec:system_model} describes the system model. Section~\ref{sec:formulation} formulates the power minimization problem under QoS constraints. Section~\ref{sec:precoding_design} presents the proposed PBB algorithm. Section~\ref{sec:hr_ris} formulates the problems based on the approximation for high-resolution RIS and presents the proposed BM algorithm. 
Section~\ref{sec:csi} expands the formulations of Sections~\ref{sec:formulation} and \ref{sec:hr_ris} for the case of imperfect CSI. Section~\ref{sec:numerical_results_ris} discusses numerical results and Section \ref{sec:conclusion} gives the conclusions.
Regarding the notation, bold lowercase and uppercase letters indicate vectors and matrices, respectively. Non-bold letters express scalars. The operators $(\cdot)^*$, $(\cdot)^T$, and $(\cdot)^H$ denote complex conjugation, transposition, and hermitian, respectively. The $i$-th element of a given vector $\boldsymbol{a}$ is denoted by $\pr{\boldsymbol{a}}_i$. 
The operator $R(\cdot)$ converts a complex-valued vector into a specific equivalent real-valued notation. For a given column vector $\boldsymbol{a} \in \mathbb{C}^N$ the equivalent real-valued vector $\boldsymbol{a}_\text{r}=R(\boldsymbol{a})$ reads as $\boldsymbol{a}_{\text{r}}=		\begin{bmatrix} \mathrm{Re} \left\{\pr{\boldsymbol{a}}_1\right\} \   \mathrm{Im} \left\{\pr{\boldsymbol{a}}_1\right\} \
\cdots \
\mathrm{Re} \left\{\pr{\boldsymbol{a}}_N\right\} \  \mathrm{Im} \left\{\pr{\boldsymbol{a}}_N\right\}
\end{bmatrix}^T$.
The operator $C(\cdot)$ converts equivalent real-valued notation into complex-valued notation. For a given matrix $\boldsymbol{A}$, $\pr{\boldsymbol{A}}_{i,j}$ denotes the element of the $i$-th row and $j$-th column and $\pr{\boldsymbol{A}}_{(i,:)}$ denotes $i$-th row of $\boldsymbol{A}$.
Finally, for the given vectors $\boldsymbol{a}$ and $\boldsymbol{b}$, $\text{P}(\boldsymbol{a}=\boldsymbol{b})$ denotes the probability of the event $\boldsymbol{a}=\boldsymbol{b}$.

\section{System Model}
\label{sec:system_model}
\begin{figure}
\captionsetup{justification=centering}
\centering
\input{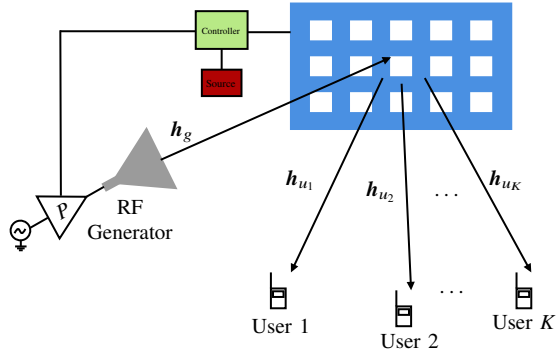}
\caption{MU-MIMO downlink realized via the RIS-based transmitter}
\label{fig:system_model}       
\end{figure}
\textcolor{r1}{The system model, shown in Fig.~\ref{fig:system_model}, consists of a MIMO transmitter realized with an RF generator illuminating a RIS with $N$ reflecting elements that serve $K$ single antenna users, similarly as in \cite{liu2021intelligent}. A symbol-level transmission is considered where the data symbol of the $k$-th user is denoted as $s_k$. For all $k\in\mathcal{K}=\chav{1,\hdots,K}$, $s_k\in \mathcal{S}$, where $\mathcal{S}$ represents all possible symbols of a $\alpha_{s}$-PSK modulation.} The symbols of all users are described in a stacked vector notation as $\boldsymbol{s}=[{s}_1,\ldots,{s}_K]^T$. Based on $\boldsymbol{s}$ the controller determines the phase shift vector $\boldsymbol{\theta}=\pr{\theta_1,\hdots,\theta_N}^T$, where $\theta_n$, with $n\in\mathcal{N}=\chav{1,\hdots,N}$, is considered to belong to the set $\mathcal{T}$ which is given by $\mathcal{T}=\{\theta: \theta= e^  \frac{j\pi (2\pi i+1) }{\alpha_{\theta}}  \textrm{,  for  }  i=1,\ldots, \alpha_{\theta} \}$ and the power, $P$, of the unmodulated carrier signal emitted by the RF generator. It is considered that each user has a SEP requisite, $\rho_k$ with $k\in\mathcal{K}$, which should be attained for the proper function of the user's application. 
With this, the received signal of the $k$-th user $z_k$, for all $k\in \mathcal{K}$, reads as ${z}_k= \sqrt{P}\boldsymbol{h}_{k}^H\boldsymbol{\theta}+{w}_k$, 
with ${w}_k\sim \mathcal{CN}({0},\sigma_w^2)$ representing additive white Gaussian noise, and, $\boldsymbol{h}_k=\boldsymbol{h}_{u_k}^H\text{diag}\pc{\boldsymbol{h}_{g}}$, being the effective channel, where $\boldsymbol{h}_{g}$ and $\boldsymbol{h}_{u_k}$ denote the channels between the RF generator and the RIS and the $k$-th user and the RIS, respectively. \textcolor{r1}{Note that, the received signal $z_k$ implicitly contains interference of other users due to $\boldsymbol{\theta}$ being dependent on $s_k$ for $k\in \mathcal{K}$. Finally, at the user's terminals, the received symbols are hard detected, which yields the detected symbol vector $\hat{\boldsymbol{s}}=\pr{\hat{s}_1, \hdots, \hat{s}_K}$.}

\section{Problem formulation}
\label{sec:formulation}

\textcolor{r1}{Considering hard detection the received symbol is $z_k$ is detected as $s_i$ if $z_k \in \mathcal{S}_i$, where $\mathcal{S}_i$ denotes the decision region of $s_i$, i.e., set of points closer to $s_i$ than all other valid candidates for detection. For PSK the decision regions are circle sectors with infinite radius and angle of $2\phi$, where $\phi=\sfrac{\pi}{\alpha_s}$. With this, the SEP of the $k$-th user \textcolor{r3}{is} written as 
\begin{align}
\label{eq:prob_gaussian_ris}
    \text{P}\pc{\hat{s}_k\neq{s}_k|\boldsymbol{\theta},P}&=1-\text{P}\pc{z_k \in \mathcal{S}_k|\boldsymbol{\theta},P}\notag\\
    &=1-\frac{1}{\pi \sigma_w^2} \int_{\mathcal{S}_k} \hspace{-0.0em} \text{e}^{-\frac{\PM{r-\sqrt{P}\boldsymbol{h}_{k}^H\boldsymbol{\theta}}^2}{\sigma_w^2}} d{r}.
\end{align}
Although, for the general case of $M$-PSK users' data, the integral in \eqref{eq:prob_gaussian_ris} requires computation via Monte-Carlo methods, a solution with tabled functions can be cast for $\alpha_s\in\{2,4\}$. In such cases, real and imaginary parts can be considered independent. This allows the decision region $\mathcal{S}_k$ to be decomposed as $\mathcal{R}_k \cap \mathcal{I}_k$, where $\mathcal{R}_k$ and $\mathcal{I}_k$ are the decision regions of the real and imaginary parts of $s_k$. \textcolor{r1}{With this, the SEP for a given RIS' reflecting coefficient vector and transmit power can be expressed as $\text{P}\pc{\hat{s}_k\neq{s}_k|\boldsymbol{\theta},P}= 1- \text{P}\pc{\hat{s}_k={s}_k|\boldsymbol{\theta},P}= 1- \text{P}\pc{z_k \in \mathcal{R}_k|\boldsymbol{\theta},P}\text{P}\pc{z_k \in \mathcal{I}_k|\boldsymbol{\theta},P}$,} with
\begin{align}
    \text{P}\pc{z_k \in \mathcal{R}_k|\boldsymbol{\theta},P}&=\int_{0}^\infty \frac{1}{\sqrt{\pi \sigma_w^2}}  e^{-\frac{P\pc{t-\text{sign}(\text{Re}\chav{s_k}) \text{Re}\{\boldsymbol{h}_k^H \boldsymbol{\theta}\}}^2}{\sigma_w^2}} dt\notag\\
    &=\Phi\pc{\sqrt{\sfrac{P}{\sigma_w^2}}v_{1,k}(\boldsymbol{\theta}) },\\
    \text{P}\pc{z_k \in \mathcal{I}_k|\boldsymbol{\theta},P}&=\int_{0}^\infty \frac{1}{\sqrt{\pi \sigma_w^2}}\  e^{-\frac{P\pc{t-\text{sign}(\text{Im}\chav{s_k}) \text{Im}\{\boldsymbol{h}_k^H \boldsymbol{\theta}\}}^2}{\sigma_w^2}} dt\notag\\
    &=\Phi\pc{\sqrt{\sfrac{P}{\sigma_w^2}}v_{2,k}(\boldsymbol{\theta}) },
\end{align}
where $\Phi(\cdot)$ denotes the cumulative Gaussian distribution function, $ v_{1,k}\pc{\boldsymbol{\theta}}=r_k \text{Re}\{\boldsymbol{h}_k^H \boldsymbol{\theta}\}$, $v_{2,k}(\boldsymbol{\theta})= i_k \text{Im}\{\boldsymbol{h}_k^H \boldsymbol{\theta}\}$, with $r_k=\sqrt{2}\ \text{sign}\pc{\text{Re}\chav{s_k}}$ and $i_k=\sqrt{2}\ \text{sign}\pc{\text{Im}\chav{s_k}}$. As a consequence, the SEP of the $k$-th user reads as 
\begin{align}
\label{eq:sep}
    \text{P}\pc{\hat{s}_k\neq{s}_k|\boldsymbol{\theta},P}=1-\Phi\pc{\sqrt{\frac{P}{\sigma_w^2}}v_{1,k}(\boldsymbol{\theta}) }\Phi\pc{\sqrt{\frac{P}{\sigma_w^2}}v_{2,k}(\boldsymbol{\theta}) }.
\end{align}
As mentioned \eqref{eq:sep} is only valid for $\alpha_s\in\{2,4\}$ and, for other PSK cases, computing the SEP requires Monte-Carlo methods. Yet, for the general $M$-PSK case, it is possible to compute the UBSEP as an upper bound on the SEP written with tabled functions. The union-bound inequality states that for any finite set of events, $\text{P} \left(\bigcup _{i}A_{i}\right)\leq \sum _{i}{\text{P} }(A_{i})$, with $A_i$ representing an event.
With this, $\text{P}\pc{\hat{s}_k\neq s_k|\boldsymbol{\theta},P}$ is bounded by 
\begin{align}
\label{eq:union_bound}
        \text{P}\pc{\hat{s}_k\neq s_k|\boldsymbol{\theta},P}& =\text{P}\pc{z_k \in \mathcal{Z}_1 \cup\mathcal{Z}_2 |\boldsymbol{\theta},P} \\
        &\leq \text{P}\pc{z_k \in \mathcal{Z}_1|\boldsymbol{\theta},P} + \text{P}\pc{z_k \in \mathcal{Z}_2|\boldsymbol{\theta},P} \notag\\
        &= \text{P}_\text{ub}\pc{\hat{s}_k|\boldsymbol{\theta},P}, \notag
\end{align}
with $\mathcal{Z}_1$ and $\mathcal{Z}_2$ depicted in Fig.~\ref{fig:union_bound_ris}.
\begin{figure}[t] 
\centering
\tikzset{every picture/.style={line width=0.75pt}} 

\begin{tikzpicture}[x=0.2pt,y=0.2pt,yscale=-1,xscale=1]

\draw    (236,421) -- (561,95) ;
\draw    (235.13,95.41) -- (559.13,418.41) ;
\draw  (152,257.41) -- (644,257.41)(398.13,23) -- (398.13,493) (637,252.41) -- (644,257.41) -- (637,262.41) (393.13,30) -- (398.13,23) -- (403.13,30)  ;
\draw  [color={rgb, 255:red, 65; green, 117; blue, 5 }  ,draw opacity=1 ][fill={rgb, 255:red, 65; green, 117; blue, 5 }  ,fill opacity=1 ] (438.75,142) -- (442.42,142) -- (442.42,144.75) -- (445.17,144.75) -- (445.17,148.58) -- (442.42,148.58) -- (442.42,151.33) -- (438.75,151.33) -- (438.75,148.58) -- (436,148.58) -- (436,144.75) -- (438.75,144.75) -- cycle ;
\draw  [color={rgb, 255:red, 255; green, 0; blue, 0 }  ,draw opacity=1 ][fill={rgb, 255:red, 252; green, 0; blue, 0 }  ,fill opacity=1 ] (507.58,210) -- (511.25,210) -- (511.25,212.75) -- (514,212.75) -- (514,216.58) -- (511.25,216.58) -- (511.25,219.33) -- (507.58,219.33) -- (507.58,216.58) -- (504.83,216.58) -- (504.83,212.75) -- (507.58,212.75) -- cycle ;
\draw  [color={rgb, 255:red, 65; green, 117; blue, 5 }  ,draw opacity=1 ][fill={rgb, 255:red, 65; green, 117; blue, 5 }  ,fill opacity=1 ] (506.75,296.67) -- (510.42,296.67) -- (510.42,299.42) -- (513.17,299.42) -- (513.17,303.25) -- (510.42,303.25) -- (510.42,306) -- (506.75,306) -- (506.75,303.25) -- (504,303.25) -- (504,299.42) -- (506.75,299.42) -- cycle ;
\draw  [color={rgb, 255:red, 0; green, 0; blue, 0 }  ,draw opacity=1 ][fill={rgb, 255:red, 0; green, 0; blue, 0 }  ,fill opacity=1 ] (541.58,182) -- (545.25,182) -- (545.25,184.75) -- (548,184.75) -- (548,188.58) -- (545.25,188.58) -- (545.25,191.33) -- (541.58,191.33) -- (541.58,188.58) -- (538.83,188.58) -- (538.83,184.75) -- (541.58,184.75) -- cycle ;
\draw    (506.47,152.36) -- (541.95,185.31) ;
\draw [shift={(543.42,186.67)}, rotate = 222.87] [color={rgb, 255:red, 0; green, 0; blue, 0 }  ][line width=0.75]    (10.93,-3.29) .. controls (6.95,-1.4) and (3.31,-0.3) .. (0,0) .. controls (3.31,0.3) and (6.95,1.4) .. (10.93,3.29)   ;
\draw [shift={(505,151)}, rotate = 42.87] [color={rgb, 255:red, 0; green, 0; blue, 0 }  ][line width=0.75]    (10.93,-3.29) .. controls (6.95,-1.4) and (3.31,-0.3) .. (0,0) .. controls (3.31,0.3) and (6.95,1.4) .. (10.93,3.29)   ;
\draw    (543.43,188.67) -- (543.98,256) ;
\draw [shift={(544,258)}, rotate = 269.53] [color={rgb, 255:red, 0; green, 0; blue, 0 }  ][line width=0.75]    (10.93,-3.29) .. controls (6.95,-1.4) and (3.31,-0.3) .. (0,0) .. controls (3.31,0.3) and (6.95,1.4) .. (10.93,3.29)   ;
\draw [shift={(543.42,186.67)}, rotate = 89.53] [color={rgb, 255:red, 0; green, 0; blue, 0 }  ][line width=0.75]    (10.93,-3.29) .. controls (6.95,-1.4) and (3.31,-0.3) .. (0,0) .. controls (3.31,0.3) and (6.95,1.4) .. (10.93,3.29)   ;
\draw  [draw opacity=0][fill={rgb, 255:red, 126; green, 211; blue, 33 }  ,fill opacity=0.34 ] (662.28,256.12) .. controls (662.28,256.12) and (662.28,256.12) .. (662.28,256.12) .. controls (662.28,256.12) and (662.28,256.12) .. (662.28,256.12) .. controls (662.28,389.33) and (544.35,497.32) .. (398.89,497.32) .. controls (330.55,497.32) and (268.29,473.49) .. (221.48,434.4) -- (398.89,256.12) -- cycle ;
\draw  [draw opacity=0][fill={rgb, 255:red, 27; green, 101; blue, 196 }  ,fill opacity=0.38 ] (135.11,258) .. controls (135.11,258) and (135.11,258) .. (135.11,258) .. controls (135.11,258) and (135.11,258) .. (135.11,258) .. controls (135.11,124.79) and (253.03,16.8) .. (398.5,16.8) .. controls (466.84,16.8) and (529.1,40.63) .. (575.91,79.72) -- (398.5,258) -- cycle ;
\draw  [draw opacity=0][fill={rgb, 255:red, 255; green, 0; blue, 0 }  ,fill opacity=0.1 ] (222.06,434.87) .. controls (168.07,391.42) and (134.11,328.37) .. (134.11,258.2) .. controls (134.11,257.93) and (134.11,257.65) .. (134.11,257.37) -- (398.78,258.2) -- cycle ;
\draw  [draw opacity=0][fill={rgb, 255:red, 80; green, 227; blue, 194 }  ,fill opacity=0.74 ] (223.05,435.5) .. controls (169.07,392.05) and (135.11,329) .. (135.11,258.83) .. controls (135.11,258.55) and (135.11,258.28) .. (135.11,258) -- (399.78,258.83) -- cycle ;

\draw (486,208.4) node  [scale=0.7] [align=left]  {$s_{i}$};
\draw (450,124.4) node  [scale=0.7] [align=left]  {$s_{i-1}$};
\draw (515.17,322.82) node  [scale=0.7] [align=left]  {$s_{i+1}$};
\draw (545,154.4) node  [scale=0.7] [align=left]  {$d_{1}$};
\draw (565,220.4) node  [scale=0.7] [align=left]  {$d_{2}$};
\draw (319,104.4) node  [scale=1] [align=left]  {$\mathcal{Z}_1$};
\draw (443,382.4) node  [scale=1] [align=left]  {$\mathcal{Z}_2$};
\draw (235,306.4) node  [scale=1] [align=left]  {$\mathcal{Z}_1 \cap \mathcal{Z}_2$};

\end{tikzpicture}
\caption{\textcolor{r1}{Representation of the sets $\mathcal{Z}_1$ and $\mathcal{Z}_2$ of the $k$-th user}}
\label{fig:union_bound_ris}       
\end{figure}
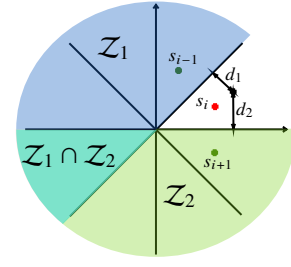
The individual probabilities are computed based on the MDDTs, $d_{1,k}$ and $d_{2,k}$, as 
\begin{align}
    \text{P}\pc{z_k \in \mathcal{Z}_1|\boldsymbol{\theta},P}&=\int_{d_{1,k}}^\infty \frac{1}{\sqrt{\pi \sigma_w^2}}\  e^{-\frac{t^2}{\sigma_w^2}} dt=\frac{1}{2}\text{erfc}\pc{\frac{d_{1,k}}{\sigma_w}}\notag\\
    \text{P}\pc{z_k \in \mathcal{Z}_2|\boldsymbol{\theta},P}&=\int_{d_{2,k}}^\infty \frac{1}{\sqrt{\pi \sigma_w^2}}\  e^{-\frac{t^2}{\sigma_w^2}} dt
    =\frac{1}{2}\text{erfc}\pc{\frac{  d_{2,k}}{\sigma_w}}.\notag
\end{align}
The MDDTs are computed by rotating the coordinate system such that the symbol of interest is placed on the real axis. 
This is done by multiplying both $s_k$ and $\boldsymbol{h}_k^H \boldsymbol{\theta}$ by $s_k^*$ which results in $s_k^* s_k=1$ and $\omega_k= s_k^* \boldsymbol{h}_k^H \boldsymbol{\theta}$.
With the rotated coordinate system the MDDTs are computed as
\begin{align}
\label{eq:d1_ris}
    d_{1,k}&=\sqrt{{P}}\pc{\text{Re}\{s_k^*\boldsymbol{h}_k^H \boldsymbol{\theta}\} \sin{\phi}-\text{Im}\{s_k^* \boldsymbol{h}_k^H \boldsymbol{\theta}\} \cos{\phi}}\\
\label{eq:d2_ris}
    d_{2,k}&=\sqrt{P}\pc{\text{Re}\{s_k^* \boldsymbol{h}_k^H \boldsymbol{\theta}\} \sin{\phi}+\text{Im}\{s_k^*\boldsymbol{h}_k^H \boldsymbol{\theta}\} \cos{\phi}}.
\end{align}
With this, the bound on $\text{P}\pc{\hat{s}_k\neq s_k|\boldsymbol{\theta},P}$ is given by
\begin{align}
\label{eq:ubsep}
    \text{P}\pc{\hat{s}_k\neq s_k|\boldsymbol{\theta},P}& \leq\text{P}_\text{ub}\pc{\hat{s}_k|\boldsymbol{\theta},P}\\
    &=\frac{1}{2}\text{erfc}\pc{\frac{d_{1,k}\pc{\boldsymbol{\theta},P}}{\sigma_w}} +\frac{1}{2}\text{erfc}\pc{\frac{d_{2,k}\pc{\boldsymbol{\theta},P}}{\sigma_w}}. \notag
\end{align}
}

\subsection{\textcolor{r1}{Formulation of the Power Minimization Problems for the General Case of Discrete Phase-Shift RIS}}
\label{sec:general_formulation}
The problem of power minimization under the condition that the SEP of each user $k$ is below the given requisite, $\rho_k$, is cast as 
\begin{align}
\label{opt:slp_original_0}
    &\min_{\boldsymbol{\theta}\in \mathcal{T}^N, P\in \mathbb{R}_+}  \ P \\
    &\hspace{0.5em}\text{s.t.}\hspace{1em} \text{P}\pc{\hat{s}_k\neq s_k|\boldsymbol{\theta},P} \leq \rho_k, \ \text{for} \ k\in \mathcal{K}. \notag
\end{align}
\textcolor{r1}{The general problem is formulated by substituting \eqref{eq:prob_gaussian_ris} in \eqref{opt:slp_original_0} which yields
 \begin{align}
    \label{opt:psep_mpsk}
    &\min_{\boldsymbol{\theta}\in \mathcal{T}^N, P\in \mathbb{R}_+}  \ P \\
    &\hspace{0.5em}\text{s.t.}\hspace{1em}1-\frac{1}{\pi \sigma_w^2} \int_{\mathcal{S}_k} \hspace{-0.0em} \text{e}^{-\frac{\PM{r-\sqrt{P}\boldsymbol{h}_{k}^H\boldsymbol{\theta}}^2}{\sigma_w^2}} d{r} \leq \rho_k, \ \text{for} \ k\in \mathcal{K}. \notag
\end{align}
As mentioned the integral has tabled solutions for $\alpha_s\in \chav{2,4}$ and for $\alpha_s\notin \chav{2,4}$ requires solution via Monte-Carlo methods. With this, for the case of $\alpha_s\in \chav{2,4}$ the exact computation of the SEP is considered. For other PSK cases solving \eqref{opt:psep_mpsk} with standard optimization problem algorithms would require the evaluation of \eqref{eq:prob_gaussian_ris} many times per iteration, which could yield prohibitive computational complexity. To avoid that, this study substitutes $\text{P}\pc{\hat{s}_k\neq {s}_k|\boldsymbol{\theta},P}$ in \eqref{opt:slp_original_0} by the UBSEP \eqref{eq:ubsep}.
}

\subsubsection{\textcolor{r1}{PSEP Problem for BPSK and QPSK users' data}}
\label{sec:psep}
For $\alpha_s\in\{2,4\}$, the power minimization problem is written by substituting \eqref{eq:sep} in \eqref{opt:slp_original_0}, which yields
\begin{align}
\label{opt:pm_sep_complex}
    &\min_{\boldsymbol{\theta}\in \mathcal{T}^N, P\in \mathbb{R}_+}  \ P \\
    &\text{s.t.} \ 1-\Phi\pc{\sqrt{\frac{P}{\sigma_w^2}}v_{1,k}(\boldsymbol{\theta}) }\Phi\pc{\sqrt{\frac{P}{\sigma_w^2}}v_{2,k}(\boldsymbol{\theta}) } \leq \rho_k, \ \text{for} \ k\in \mathcal{K}. \notag
\end{align}
Constraining the SEP to be smaller or equal to $\rho_k$ is equivalent to constraining the correct detection probability to be greater or equal to $1-\rho_k$. The PSEP problem is cast by rewriting \eqref{opt:pm_sep_complex} with real-valued variables and taking the logarithm of the correct detection probability constraints, which yields 
\begin{align}
\label{opt:pm_sep}
    &\min_{\boldsymbol{\theta}_\text{r}, P\in \mathbb{R}_+}  \ P \\
    &\hspace{0.5em}\text{s.t.}\hspace{1em} \pr{\boldsymbol{\theta}_{\text{r}}}_{2n-1}+j\pr{\boldsymbol{\theta}_{\text{r}}}_{2n} \in \mathcal{T}, \text{for} \ n \in \mathcal{N},\notag\\
    &-\sum_{\xi=1}^2\ln\pc{\Phi\pc{\sqrt{\frac{P}{\sigma_w^2}}\boldsymbol{h}_{\xi,k}^T\boldsymbol{\theta}_\text{r} }}  \leq \beta_k, \ \text{for} \ k\in \mathcal{K}, \notag
\end{align}
 where $\boldsymbol{\theta}_\text{r}=R(\boldsymbol{\theta})$, $\beta_k=-\ln(1-\rho_k)$,
 \begin{align}
     \boldsymbol{h}_{\text{1},k}^T&=\pr{[\boldsymbol{\gamma}_{R,k}]_1, -[\boldsymbol{\gamma}_{I,k}]_1,\hdots, [\boldsymbol{\gamma}_{R,k}]_N, -[\boldsymbol{\gamma}_{I,k}]_N }, \\
     \boldsymbol{h}_{\text{2},k}^T&=\pr{[\boldsymbol{\gamma}_{I,k}]_1\}, [\boldsymbol{\gamma}_{R,k}]_1,\hdots, [\boldsymbol{\gamma}_{I,k}]_N\}, [\boldsymbol{\gamma}_{R,k}]_N\}},
 \end{align}
\textcolor{r1}{with $\boldsymbol{\gamma}_{R,k}=r_k\text{Re}\{\boldsymbol{h}_k^H\}$, and, $\boldsymbol{\gamma}_{I,k}=i_k\text{Im}\{\boldsymbol{h}_k^H\}$.}


\subsubsection{\textcolor{r1}{PUBSEP Problem for the General $M$-PSK case}}
\label{sec:pubsep}
For the general $M$-PSK case the substitution of $\text{P}(\hat{s}_k\neq s_k|\boldsymbol{\theta},P)$ in \eqref{opt:slp_original_0} by $\text{P}_\text{ub}(\hat{s}_k|\boldsymbol{\theta},P)$ is considered. As the UBSEP is an upper-bound on the SEP substituting the constraint $\text{P}\pc{\hat{s}_k\neq s_k|\boldsymbol{\theta},P}\leq \rho_k$ by $\text{P}_\text{ub}\pc{\hat{s}_k|\boldsymbol{\theta},P}\leq \rho_k$ yields a restriction of the feasible set of the original problem, implying that the restricted problem's optimal transmit power is larger or equal to the original one.
Substituting \eqref{eq:ubsep} in \eqref{opt:slp_original_0} yields the following problem
\begin{align}
\label{opt:complex_form_0}
    &\min_{\boldsymbol{\theta}\in \mathcal{T}^N, P\in \mathbb{R}_+}  \ P \\
    &\hspace{0.7em}\text{s.t.}\hspace{0.5em} \sum_{\xi=1}^2 \frac{1}{2}\text{erfc}\pc{\frac{d_{\xi,k}\pc{\boldsymbol{\theta},P}}{\sigma_w}} \leq \rho_k, \ \text{for} \ k\in \mathcal{K}.\notag 
\end{align}
\textcolor{r1}{Note that, $\text{P}\pc{\hat{s}_k\neq s_k|\boldsymbol{\theta},P} \leq\text{P}_\text{ub}\pc{\hat{s}_k|\boldsymbol{\theta},P}\leq \rho_k\ll 1$. Considering $\rho_k\leq 0.5$, the $k$-th UBSEP constraint is only achievable for $d_{\xi,k}\pc{\boldsymbol{\theta},P}\geq 0$, which corresponds to having the noiseless received signal $y_k \in \mathcal{S}_k$. With this, $d_{\xi,k}\pc{\boldsymbol{\theta},P}\geq 0$, for $\xi \in \{1,2\}$ and $k\in\mathcal{K}$ is an implicit constraint of problem \eqref{opt:complex_form_0}.} The PUBSEP problem is cast writing \eqref{opt:complex_form_0} with real-valued variables and explicitly including the MDDTs restriction, which yields
\begin{align}
\label{opt:rv_slp}
    &\min_{\boldsymbol{\theta}_\text{r},P\in \mathbb{R}_+}  \ P \\
    &\hspace{0.5em}\text{s.t.}\hspace{0.5em} \pr{\boldsymbol{\theta}_\text{r}}_{2n-1}+ j\pr{\boldsymbol{\theta}_\text{r}}_{2n} \in \mathcal{T}, \ \text{for} \ n\in \mathcal{N}, \ \ \boldsymbol{U}\boldsymbol{\theta}_\text{r}\preceq \boldsymbol{0}, \notag \\
    &\hspace{0.5em} \displaystyle \sum_{\xi=1}^2 \frac{1}{2}\text{erfc}\pc{ \sqrt{\frac{P}{\sigma_w^2}}\boldsymbol{u}_{\xi,k}^T \boldsymbol{\theta}_{\text{r}} } \leq \rho_k, \ \text{for} \ k\in \mathcal{K}. \notag
\end{align}
where 
\begin{align}
\boldsymbol{u}_{1,k}&=\pc{\boldsymbol{\gamma}_{\text{R},k}\sin\pc{\phi} -\boldsymbol{\gamma}_{\text{I},k}\cos\pc{\phi}}^T, \\
\boldsymbol{u}_{2,k}&=\pc{\boldsymbol{\gamma}_{\text{R},k}\sin\pc{\phi} +\boldsymbol{\gamma}_{\text{I},k}\cos\pc{\phi}}^T,
\end{align}
with
$\boldsymbol {\gamma }_{\text {R},k}= [\textrm {Re}\{[\boldsymbol {\zeta }_{k}]_{1}\}, -\textrm {Im}\{[\boldsymbol {\zeta }_{k}]_{1}\},\ldots,\textrm {Re}\{[\boldsymbol {\zeta }_{k}]_{N}\},$ $ -\textrm {Im}\{[\boldsymbol {\zeta }_{k}]_{N}\}],$ and $\boldsymbol {\gamma }_{\text {I},k}= [\textrm {Im}\{[\boldsymbol {\zeta }_{k}]_{1}\}, \textrm {Re}\{[\boldsymbol {\zeta }_{k}]_{1}\},\ldots, $ $\textrm {Im}\{[\boldsymbol {\zeta }_{k}]_{N}\}, \textrm {Re}\{[\boldsymbol {\zeta }_{k}]_{N}\}]$ , where 
\begin{align}
\boldsymbol{\zeta}_{k}&=s_k^{*}\boldsymbol{h}_k^H, \quad \boldsymbol{U}=-[\boldsymbol{\beta}_1^T,\boldsymbol{\beta}_2^T]^T \\
\boldsymbol{\beta}_1&=[\boldsymbol{u}_{1,1}, \hdots, \boldsymbol{u}_{1,K}]^T, \quad  \boldsymbol{\beta}_2=[\boldsymbol{u}_{2,1}, \hdots, \boldsymbol{u}_{2,K}]^T.
\end{align}
 Note that, $d_{\xi,k}=\sqrt{P}\boldsymbol{u}_{\xi,k}^T\boldsymbol{\theta}_\text{r}$, for $\xi\in\{1,2\}$ and $k\in\mathcal{K}$. 



\section{Proposed Partial Branch-and-Bound Algorithm}
\label{sec:precoding_design}

\textcolor{r1}{Due to the discrete phase shift constraints, $\pr{\boldsymbol{\theta}_\text{r}}_{2n-1}+ j\pr{\boldsymbol{\theta}_\text{r}}_{2n} \in \mathcal{T}$, for $n\in \mathcal{N}$, computing the optimal solution of the PSEP and PUBSEP problems requires the utilization of FBB methods. For an introduction to FBB refer to \cite{Landau2017,lopes2021discrete,General_MMDDT_BB}. Unfortunately, FBB approaches can yield prohibitive computational complexity for some scenarios. To achieve favorable complexity performance trade-offs this study proposes a PBB method that accepts as a solution any pair $(P_\text{out}, \boldsymbol{\theta}_\text{out})\in \mathcal{H}$. The set $\mathcal{H}$ contains all feasible solutions that either attain the system's target power budget, $P_\text{b}$, or are sufficiently close to the optimal solution, {$P_\text{opt}$}, such that further computation is considered unnecessary. In mathematical terms 
$\mathcal{H}=\chav{(P, \boldsymbol{\theta}):  \boldsymbol{\theta}\in \mathcal{T}^N \wedge \pc {P\leq P_\text{b} \lor 10\log_{10}(\sfrac{P}{P_\text{opt}})\leq \gamma} }$, with $\gamma$ being the acceptable power increase factor.}

\textcolor{r1}{The first step of the proposed algorithm, consists of the computation of an upper-bound solution and the evaluation of the stopping criteria. To this end, the original problems are relaxed substituting the discrete feasible set $\mathcal{T}^N$ by convex hull $\mathcal{P}$, described as $\boldsymbol{A}\boldsymbol{\theta}_\text{r}\preceq \boldsymbol{b}$, where $\boldsymbol{A}=\begin{bmatrix} (\boldsymbol{I}_N\otimes \boldsymbol{\beta}_1)^T,  (\boldsymbol{I}_N\otimes \boldsymbol{\beta}_2)^T, \ldots, (\boldsymbol{I}_N\otimes \boldsymbol{\beta}_{\alpha_\theta})^T \end{bmatrix}^T$, $\boldsymbol{\beta}_i= [\cos{\pc{\sfrac{2\pi i }{\alpha_\theta}}}, \ -\sin{\pc{\sfrac{2\pi i }{\alpha_\theta}}}]$, for $i\in\chav{1,\ldots,  \alpha_\theta}$, and, $\boldsymbol{b}={\cos\pc{\sfrac{\pi}{\alpha_\theta}}}\boldsymbol{1}_{N\alpha_\theta}$,
with $\boldsymbol{1}_{N\alpha_\theta}$ being the column vector with $N\alpha_\theta$ ones. By substituting $\mathcal{T}^N$ by its convex hull, the optimization problems described in \eqref{opt:pm_sep} and \eqref{opt:rv_slp} are cast in the form of 
\begin{align}
\label{opt:rv_slp_relaxed}
    &\min_{\boldsymbol{\theta}_\text{r},P\in \mathbb{R}_+}  \ P \\
    &\hspace{0.5em}\text{s.t.}\hspace{0.5em} f_k(\boldsymbol{\theta}_\text{r},P) \leq 0, \ \text{for} \ k\in \mathcal{K}, \ \boldsymbol{G}\boldsymbol{\theta}_\text{r}\preceq \boldsymbol{t}, \notag
\end{align}
where, for the PSEP case, $\boldsymbol{G}=\boldsymbol{A}$, $ \boldsymbol{t}=\boldsymbol{b}$ and $f_k(\boldsymbol{\theta}_\text{r},P)=-\sum_{\xi=1}^2\ln\pc{\Phi\pc{\sqrt{\frac{P}{\sigma_w^2}}\boldsymbol{h}_{\xi,k}^T\boldsymbol{\theta}_\text{r} }} - \beta_k$, and for the PUBSEP case, $\boldsymbol{G}=[\boldsymbol{A}^T, \boldsymbol{U}^T]^T $ and $ \boldsymbol{t}=[\boldsymbol{b}^T, \boldsymbol{0}^T]^T$, and $f_k(\boldsymbol{\theta}_\text{r},P)=\sum_{\xi=1}^2 \frac{1}{2}\text{erfc}\pc{ \sqrt{\frac{P}{\sigma_w^2}}\boldsymbol{u}_{\xi,k}^T \boldsymbol{\theta}_{\text{r}} }-\rho_k$. \textcolor{r2}{In general, solving a relaxed problem also serves as an infeasibility test of the original problem. With this, if \eqref{opt:rv_slp_relaxed} is infeasible, the original problem \eqref{opt:slp_original_0} is also infeasible.} By optimally solving \eqref{opt:rv_slp_relaxed} one gets $\boldsymbol{\theta}_\text{r,lb}$, which implies $\boldsymbol{\theta}_\text{lb}=C(\boldsymbol{\theta}_\text{r,lb})\in \mathcal{P}$, and its corresponding transmit power ${P}_\text{lb}$. Note that, $\boldsymbol{\theta}_\text{lb}\in \mathcal{P}$ can also belong to $\mathcal{T}^N$ once $\mathcal{P}\cap\mathcal{T}^N\neq \emptyset$. If this is the case, $P_\text{opt}={P}_\text{lb}$ and $\boldsymbol{\theta}_\text{opt}=\boldsymbol{\theta}_\text{lb}$, with $P_\text{opt}$ and $\boldsymbol{\theta}_\text{opt}$ being the optimal solutions of the original problem. Yet, if $\boldsymbol{\theta}_\text{lb}\notin \mathcal{T}^N$ an upper bound solution on $\boldsymbol{\theta}_\text{opt}$ is achieved by projecting to $\boldsymbol{\theta}_\text{lb}$ to $\mathcal{T}^N$. The projection method considered is uniform quantization (UQ), denoted by the operator $Q(\cdot)$. By this approach, the $p$-th entry of $\boldsymbol{\theta}_\text{ub}=Q(\boldsymbol{\theta}_\text{lb})$, denoted as $\theta_{\text{ub},p}$, is computed as $\theta_{\text{ub},p}=\displaystyle \text{arg} \hspace{-1em}\min_{i\in \{1,\hdots,\alpha_\theta\}} |\theta_{\text{lb},p}-\theta_i|$, where $\theta_{\text{lb},p}$ denotes the $p$-th entry of $\boldsymbol{\theta}_\text{lb}$ and $\theta_i$ is the $i$-th element of $\mathcal{T}$. Based on $\boldsymbol{\theta}_{\text{r,ub}}=R(\boldsymbol{\theta}_{\text{ub}})$, the  transmit power is given by the solution of the following univariate upper bounding problem
\begin{align}
\label{opt:ub_power}
&P_\text{ub}= \min_{P\in \mathbb{R}_+}  \ P \\
   &\hspace{0.5em}\text{s.t.}\hspace{0.5em} f_k(\boldsymbol{\theta}_{\text{r,ub}},P) \leq 0, \ \text{for} \ k\in \mathcal{K}. \notag
\end{align}
If $P_\text{ub}\leq P_\text{b}$, the upper bound solution pair ($P_\text{ub}$, $\boldsymbol{\theta}_{\text{ub}}$) attains the target power budget of the system. On the other hand, if $10\log_{10} \pc{\sfrac{P_\text{ub}}{P_\text{lb}}} \leq\gamma$ the solution pair is sufficiently close to the optimal solution such that further computation is considered unnecessary. With this, for both cases, the algorithm terminates with $\boldsymbol{\theta}_{\text{out}}=\boldsymbol{\theta}_{\text{ub}}$ and $P_\text{out}=P_\text{ub}$. In what follows, the PSEP and PUBSEP concepts are applied to \eqref{opt:rv_slp_relaxed} and \eqref{opt:ub_power}.
By substituting $\mathcal{T}^N$ by its convex hull, the relaxed PSEP problem is cast as  
\begin{align}
\label{opt:rv_slp_relaxed_sep}
    &\min_{\boldsymbol{\theta}_\text{r},P }  \ P \\
    &\hspace{0.5em}\text{s.t.}\hspace{0.5em} \boldsymbol{A}\boldsymbol{\theta}_\text{r}\preceq \boldsymbol{b},  \ P\geq 0, \ \notag \\
   &-\sum_{\xi=1}^2\ln\pc{\Phi\pc{\sqrt{\frac{P}{\sigma_w^2}}\boldsymbol{h}_{\xi,k}^T\boldsymbol{\theta}_\text{r} }} \leq \beta_k, \ \text{for} \ k\in \mathcal{K}. \notag
\end{align}
An equivalent problem is formulated by defining the vector $\boldsymbol{x}=[\sqrt{P} \boldsymbol{\theta}_\text{r}^T,\sqrt{P}]^T$ and applying the square root to the objective, which reads as 
\begin{align}
\label{opt:power_min_sep}
    &\min_{\boldsymbol{x}}  \ \boldsymbol{c}^T\boldsymbol{x} \\
    &\hspace{0.5em}\text{s.t.}\hspace{0.5em} \boldsymbol{R}\boldsymbol{x}\preceq \boldsymbol{0}_{(N\alpha_\theta+1)}, \notag\\
    &\hspace{0.5em} -\sum_{\xi=1}^2 \ln\pc{\Phi\pc{\boldsymbol{q}_{\xi,k}^T \boldsymbol{x} }}  \leq \beta_k, \ \text{for} \ k\in \mathcal{K}, \notag
\end{align}
where $\boldsymbol{q}_{1,k}=(\sfrac{1}{\sigma_w})[\boldsymbol{h}_{1,k}^T, 0]^T$, $\boldsymbol{q}_{2,k}=(\sfrac{1}{\sigma_w})[\boldsymbol{h}_{2,k}^T, 0]^T$, $\boldsymbol{c}=[\boldsymbol{0}_{2N}^T,1]^T$, $\boldsymbol{M}=[\boldsymbol{A}, \boldsymbol{0}_{N\alpha_\theta}]$, 
$\boldsymbol{R}=[(\boldsymbol{M}-\boldsymbol{b}\boldsymbol{c}^T)^T, -\boldsymbol{c} ]^T$. As demonstrated in Appendix \ref{app:sep}, the SEP constraint functions from \eqref{opt:power_min_sep} are convex in $\boldsymbol{x}$. With this, \eqref{opt:power_min_sep} is a convex problem solvable with the barrier method \cite[Section 11.3]{Boyd_2004}. From the solution of \eqref{opt:power_min_sep}, termed $\boldsymbol{x}_\text{lb}$, one can extract ${P}_\text{lb}$ and $\boldsymbol{\theta}_\text{lb}=C(\boldsymbol{\theta}_\text{r,lb})$. The upper bound solution $\boldsymbol{\theta}_\text{ub}=Q(\boldsymbol{\theta}_\text{lb})$ can be converted to real-valued notation as $\boldsymbol{\theta}_\text{r,ub}=R(\boldsymbol{\theta}_\text{ub})$ and utilized for computing the upper bound transmit power $P_\text{ub}$ with
\begin{align}
\label{opt:ub_power_sep}
P_\text{ub}=& \min_{P\in \mathbb{R}_+}  \ P \\
   &\text{s.t.}\hspace{0.5em} -\sum_{\xi=1}^2\ln\pc{\Phi\pc{\sqrt{\frac{P}{\sigma_w^2}}\boldsymbol{h}_{\xi,k}^T\boldsymbol{\theta}_\text{r,ub} }} \leq \beta_k, \ \text{for} \ k\in \mathcal{K}. \notag
\end{align}
If either $\boldsymbol{h}_{1,k}^T \boldsymbol{\theta}_{\text{r,ub}}\leq 0$ or $\boldsymbol{h}_{2,k}^T \boldsymbol{\theta}_{\text{r,ub}}\leq 0$ for any $k \in \mathcal{K}$, problem \eqref{opt:ub_power_sep} is infeasible for $\rho_k<0.5$ and $P_\text{ub}=\infty$.
As in the PSEP case, the PUBSEP relaxed problem is formulated with $\boldsymbol{x}=[\sqrt{P} \boldsymbol{\theta}_\text{r}^T,\sqrt{P}]^T$ as 
\begin{align}
\label{opt:power_min_ubsep}
    &\min_{\boldsymbol{x}}  \ \boldsymbol{c}^T\boldsymbol{x} \\
    &\hspace{0.5em}\text{s.t.}\hspace{0.5em} \boldsymbol{D}\boldsymbol{x}\preceq \boldsymbol{0}_{(N \alpha_\theta +2K+1)}, \notag\\
    &\hspace{0.5em} \frac{1}{2}\sum_{\xi=1}^2 \text{erfc}\pc{ \boldsymbol{\nu}_{\xi,k}^T \boldsymbol{x} } \leq \rho_k, \ \text{for} \ k\in \mathcal{K}, \notag
\end{align}
where $\boldsymbol{\nu}_{1,k}=(\sfrac{1}{\sigma_w})[\boldsymbol{u}_{1,k}^T, 0]^T$, $\boldsymbol{\nu}_{2,k}=(\sfrac{1}{\sigma_w})[\boldsymbol{u}_{2,k}^T, 0]^T$, $\boldsymbol{D}=[\boldsymbol{R}^T, \boldsymbol{C}^T, -\boldsymbol{c}]^T$ and $\boldsymbol{C}=[\boldsymbol{U}, \boldsymbol{0}]$.
\textcolor{r2}{As discussed in Appendix \ref{app:ubsep}, the UBSEP constraints under the condition $\boldsymbol{C}\boldsymbol{x}\preceq\boldsymbol{0}$ are convex, which implies that \eqref{opt:power_min_ubsep} is a convex problem solvable with the barrier method.} 
From $\boldsymbol{x}_\text{lb}$ one can extract ${P}_\text{lb}$ and $\boldsymbol{\theta}_\text{lb}=C(\boldsymbol{\theta}_\text{r,lb})$. The upper bound solution $\boldsymbol{\theta}_\text{ub}=Q(\boldsymbol{\theta}_\text{lb})$ can be converted to real-valued notation as $\boldsymbol{\theta}_\text{r,ub}=R(\boldsymbol{\theta}_\text{ub})$ and utilized for computing the upper bound transmit power $P_\text{ub}$ with
\begin{align}
\label{opt:ub_power_ubsep}
P_\text{ub}&= \min_{P\in \mathbb{R}_+}  \ P \\
   &\hspace{0.5em}\text{s.t.}\hspace{0.5em} \frac{1}{2}\sum_{\xi=1}^2 \text{erfc}\pc{ \sqrt{\frac{P}{\sigma_w^2}}\boldsymbol{u}_{\xi,k}^T \boldsymbol{\theta}_{\text{r,ub}} } \leq \rho_k, \ \text{for} \ k\in \mathcal{K}. \notag
\end{align}
As before, if $\boldsymbol{u}_{\xi,k}^T \boldsymbol{\theta}_{\text{r,ub}}\leq 0$, for any $\xi\in\{1,2\}$ and $k \in \mathcal{K}$, problem \eqref{opt:ub_power_ubsep} is infeasible for $\rho_k<0.5$ and $P_\text{ub}=\infty$.
}
\subsection{Branch-and-Bound Tree Search Stage}

\begin{figure} 
\centering
\tikzset{every picture/.style={line width=0.75pt}} 

\begin{tikzpicture}[x=0.13pt,y=0.26pt,yscale=-1,xscale=1]

\draw    (736,46) -- (1288,138) ;
\draw    (736,46) -- (184,138) ;
\draw    (736,46) -- (920,138) ;
\draw    (736,46) -- (552,138) ;
\draw    (184,138) -- (322,230) ;
\draw    (184,138) -- (230,230) ;
\draw    (184,138) -- (138,230) ;
\draw    (184,138) -- (46,230) ;
\draw    (552,138) -- (690,230) ;
\draw    (552,138) -- (598,230) ;
\draw    (552,138) -- (506,230) ;
\draw    (552,138) -- (414,230) ;
\draw    (920,138) -- (1058,230) ;
\draw    (920,138) -- (966,230) ;
\draw    (920,138) -- (874,230) ;
\draw    (920,138) -- (782,230) ;
\draw    (1288,138) -- (1426,230) ;
\draw    (1288,138) -- (1334,230) ;
\draw    (1288,138) -- (1242,230) ;
\draw    (1288,138) -- (1150,230) ;
\draw  [dash pattern={on 4.5pt off 4.5pt}]  (0,138) -- (309.5,138) -- (1455.5,138) ;
\draw  (1500,142.65) -- (1735.22,142.65)(1614.89,69) -- (1614.89,207) (1728.22,137.65) -- (1735.22,142.65) -- (1728.22,147.65) (1609.89,76) -- (1614.89,69) -- (1619.89,76)  ;
\draw   (1542.29,142.65) .. controls (1542.29,119.13) and (1574.79,100.06) .. (1614.89,100.06) .. controls (1654.99,100.06) and (1687.49,119.13) .. (1687.49,142.65) .. controls (1687.49,166.18) and (1654.99,185.25) .. (1614.89,185.25) .. controls (1574.79,185.25) and (1542.29,166.18) .. (1542.29,142.65) -- cycle ;
\draw  [color={rgb, 255:red, 0; green, 0; blue, 0 }  ,draw opacity=1 ][fill={rgb, 255:red, 0; green, 0; blue, 0 }  ,fill opacity=1 ] (1657.83,107.96) .. controls (1662.37,105.3) and (1669.73,105.3) .. (1674.27,107.96) .. controls (1678.81,110.63) and (1678.81,114.94) .. (1674.27,117.6) .. controls (1669.73,120.26) and (1662.37,120.26) .. (1657.83,117.6) .. controls (1653.3,114.94) and (1653.3,110.63) .. (1657.83,107.96) -- cycle ;
\draw  [color={rgb, 255:red, 0; green, 0; blue, 0 }  ,draw opacity=1 ][fill={rgb, 255:red, 0; green, 0; blue, 0 }  ,fill opacity=1 ] (1555.16,168.2) .. controls (1559.7,165.54) and (1567.06,165.54) .. (1571.6,168.2) .. controls (1576.14,170.86) and (1576.14,175.18) .. (1571.6,177.84) .. controls (1567.06,180.5) and (1559.7,180.5) .. (1555.16,177.84) .. controls (1550.63,175.18) and (1550.63,170.86) .. (1555.16,168.2) -- cycle ;
\draw  [color={rgb, 255:red, 0; green, 0; blue, 0 }  ,draw opacity=1 ][fill={rgb, 255:red, 0; green, 0; blue, 0 }  ,fill opacity=1 ] (1555.17,107) .. controls (1559.39,104.15) and (1566.48,104) .. (1571.02,106.66) .. controls (1575.56,109.32) and (1575.81,113.79) .. (1571.6,116.64) .. controls (1567.38,119.48) and (1560.28,119.64) .. (1555.74,116.98) .. controls (1551.2,114.32) and (1550.95,109.85) .. (1555.17,107) -- cycle ;
\draw  [color={rgb, 255:red, 0; green, 0; blue, 0 }  ,draw opacity=1 ][fill={rgb, 255:red, 0; green, 0; blue, 0 }  ,fill opacity=1 ] (1657.83,168.2) .. controls (1662.37,165.54) and (1669.73,165.54) .. (1674.27,168.2) .. controls (1678.81,170.86) and (1678.81,175.18) .. (1674.27,177.84) .. controls (1669.73,180.5) and (1662.37,180.5) .. (1657.83,177.84) .. controls (1653.3,175.18) and (1653.3,170.86) .. (1657.83,168.2) -- cycle ;

\draw (150,125) node  [scale=0.7] [align=left]    {$\theta_{1}$};
\draw (530,125) node  [scale=0.7] [align=left]    {$\theta_{2}$};
\draw (940,125) node  [scale=0.7] [align=left]    {$\theta_{3}$};
\draw (1300,125) node  [scale=0.7] [align=left]    {$\theta_{4}$};
\draw (10,225) node  [scale=0.7] [align=left]    {$\theta_{1}$};
\draw (105,225) node  [scale=0.7] [align=left]    {$\theta_{2}$};
\draw (190,225) node  [scale=0.7] [align=left]    {$\theta_{3}$};
\draw (275,225) node  [scale=0.7] [align=left]    {$\theta_{4}$};
\draw (380,225) node  [scale=0.7] [align=left]    {$\theta_{1}$};
\draw (475,225) node  [scale=0.7] [align=left]    {$\theta_{2}$};
\draw (560,225) node  [scale=0.7] [align=left]    {$\theta_{3}$};
\draw (640,225) node  [scale=0.7] [align=left]    {$\theta_{4}$};
\draw (750,225) node  [scale=0.7] [align=left]    {$\theta_{1}$};
\draw (845,225) node  [scale=0.7] [align=left]    {$\theta_{2}$};
\draw (930,225) node  [scale=0.7] [align=left]    {$\theta_{3}$};
\draw (1010,225) node  [scale=0.7] [align=left]    {$\theta_{4}$};
\draw (1115,225) node  [scale=0.7] [align=left]    {$\theta_{1}$};
\draw (1210,225) node  [scale=0.7] [align=left]    {$\theta_{2}$};
\draw (1295,225) node  [scale=0.7] [align=left]    {$\theta_{3}$};
\draw (1380,225) node  [scale=0.7] [align=left]    {$\theta_{4}$};
\draw (16,74.4) node  [scale=0.7] [align=left]    {$p=1$};
\draw (16,171.4) node  [scale=0.7] [align=left]    {$p=2$};
\draw (1708.01,88.4) node [scale=0.7] [align=left]    {$\theta_{1}$};
\draw (1708.01,182.4) node [scale=0.7] [align=left]    {$\theta_{2}$};
\draw (1506.73,182.4) node [scale=0.7] [align=left]    {$\theta_{3}$};
\draw (1506.73,88.4) node [scale=0.7] [align=left]    {$\theta_{4}$};

\end{tikzpicture}
\caption{\textcolor{r1}{Tree representation of the set $\mathcal{T}^N$ for a system with $N=2$ reflecting elements and with a RIS with four available phase shifts ($\alpha_\theta=4$)}}
\label{fig:tree}       
\end{figure}
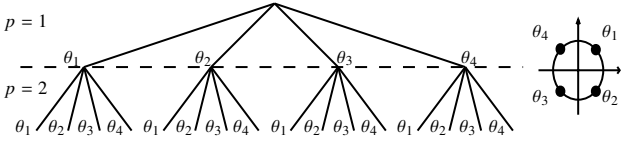

If neither stopping criteria are met at the initialization step, i.e., if $\boldsymbol{\theta}_\text{lb}\notin\mathcal{T}^N$, $P_\text{ub}>P_\text{b}$ and $10\log_{10} \pc{\sfrac{P_\text{ub}}{P_\text{lb}}} >\gamma$, the proposed PBB algorithm proceeds to the tree search stage where the tree represents the set $\mathcal{T}^N$. To this end, the smallest known upper bound is initialized as $\check{P}=P_\text{ub}$ and its reflection coefficient vector as $\check{\boldsymbol{\theta}}=\boldsymbol{\theta}_\text{ub}$. The tree is constructed considering that the $p$-th reflection coefficient represents the $p$-th layer and each possible subvector $\boldsymbol{f}\in \mathcal{T}^p$ represents one branch. 
An example of a tree for a system with $N=2$ reflecting elements and $\alpha_\theta=4$ is shown in Fig.~\ref{fig:tree}.
The tree search process performs breadth-first search to find a vector $\boldsymbol{\theta}_\text{out}\in \mathcal{T}^N$ that attains the condition $10 \log_{10}(\sfrac{P(\boldsymbol{\theta}_\text{out})}{P_\text{opt}})\leq \gamma$ with $P_\text{opt}=P(\boldsymbol{\theta}_\text{opt})$. During the search process, if an intermediate solution pair $(P_\text{int},\boldsymbol{\theta}_\text{int})$ with $\boldsymbol{\theta}_\text{int}\in \mathcal{T}^N$ and $P_\text{int}\leq P_\text{b}$ is found, the process terminates with $\boldsymbol{\theta}_\text{out}=\boldsymbol{\theta}_\text{int}$ and $P_\text{out}=P_\text{int}$.
The process starts at layer value $p=1$ by fixing $p$ entries of $\boldsymbol{\theta}$ such that the reflection coefficient vector becomes $\boldsymbol{\theta}=[\boldsymbol{f}_i^T, \boldsymbol{v}^T]^T$ with $\boldsymbol{f}_i\in \mathcal{T}^p$ and $\boldsymbol{\theta}_\text{r}=R(\boldsymbol{\theta})=[\boldsymbol{f}_{\text{r},i}^T, \boldsymbol{v}_\text{r}^T]^T$. With this, a subproblem is assembled as
\begin{align}
\label{opt:subproblem}
    \{P_{\text{opt}|\boldsymbol{f}_i}&,\boldsymbol{v}_{\text{r,opt}|\boldsymbol{f}_i}\}=\min_{\boldsymbol{v}_\text{r},P\in\mathbb{R}_+}  \ P \\
    &\hspace{-2em}\text{s.t.}\hspace{0.5em} \pr{\boldsymbol{v}_\text{r}}_{2n-1}+ j\pr{\boldsymbol{v}_\text{r}}_{2n} \in \mathcal{T}, \ \text{for} \ n\in\{1,\hdots,N-p\}, \notag \\
    & \boldsymbol{U}^\prime \boldsymbol{v}_\text{r} + \boldsymbol{U}_\text{fixed}\boldsymbol{f}_\text{r,i} \preceq \boldsymbol{0}, \ \ f_k(\boldsymbol{f}_{\text{r},i},\boldsymbol{v}_\text{r},P)\leq 0, \ \text{for} \ k\in \mathcal{K}, \notag
\end{align}
where $P_{\text{opt}|\boldsymbol{f}_i}$ is the optimal transmit power for the fixed vector $\boldsymbol{f}_i$, $\boldsymbol{U}_\text{fixed}$ and $\boldsymbol{U}^\prime$ correspond to the first $2p$ and subsequent $2(N-p)$ columns of $\boldsymbol{U}$, and, the constraint $\boldsymbol{U}^\prime \boldsymbol{v}_\text{r} \preceq \boldsymbol{U}_\text{fixed}\boldsymbol{f}_\text{r,i}$ is only taken into account for the PUBSEP case. A lower bounding subproblem on $P_{\text{opt}|\boldsymbol{f}_i}$ is obtained by relaxing $\mathcal{T}^{(N-p)}$ to its convex hull $\mathcal{J}$, which yields
\begin{align}
\label{opt:lb_subproblem_pre}
    \{P_{\text{lb}|\boldsymbol{f}_i},\boldsymbol{v}_{\text{r,lb}|\boldsymbol{f}_i}\}&=\min_{\boldsymbol{v}_\text{r},P\in\mathbb{R}_+}  \ P \\
    &\hspace{-1.5em}\text{s.t.}\hspace{0.5em}  C(\boldsymbol{v}_\text{r}) \in \mathcal{J}, \ \ \boldsymbol{U}^\prime \boldsymbol{v}_\text{r} + \boldsymbol{U}_\text{fixed}\boldsymbol{f}_\text{r,i} \preceq \boldsymbol{0}, \ \ \notag\\
    &f_k(\boldsymbol{f}_{\text{r},i},\boldsymbol{v}_\text{r},P)\leq 0, \ \text{for} \ k\in \mathcal{K}, \notag
\end{align}
where the constraint $\boldsymbol{U}^\prime \boldsymbol{v}_\text{r} \preceq \boldsymbol{U}_\text{fixed}\boldsymbol{f}_\text{r,i}$ is only taken into account for the PUBSEP case.
An upper bound on $P_{\text{opt}|\boldsymbol{f}_i}$ can be computed by projecting the vector $\boldsymbol{v}_{\text{lb}|\boldsymbol{f}_i}=C(\boldsymbol{v}_{\text{r,lb}|\boldsymbol{f}_i})$ to $\mathcal{T}^{(N-p)}$ resulting in $\boldsymbol{v}_{\text{ub}|\boldsymbol{f}_i}=Q(\boldsymbol{v}_{\text{lb}|\boldsymbol{f}_i})$ and computing the corresponding transmit power $P_{\text{ub}|\boldsymbol{f}_i}$, with $\boldsymbol{\theta}_{\text{r,ub}|\boldsymbol{f}_i}=R(\boldsymbol{\theta}_{\text{ub}|\boldsymbol{f}_i})=R([\boldsymbol{f}_i^T,\boldsymbol{v}_{\text{ub}|\boldsymbol{f}_i}^T ]^T)$ using \eqref{opt:ub_power}. If the upper bound solution $P_{\text{ub}|\boldsymbol{f}_i}\leq P_\text{b}$, the solution pair, $(P_{\text{ub}|\boldsymbol{f}_i}, \boldsymbol{\theta_{\text{ub}|\boldsymbol{f}_i}})$, attains the low-resolution constraints and the target power budget of the system. With this, the algorithm terminates with $P_\text{out}=P_{\text{ub}|\boldsymbol{f}_i}$ and $\boldsymbol{\theta}_\text{out}=\boldsymbol{\theta_{\text{ub}|\boldsymbol{f}_i}}$. If, however, $P_{\text{ub}|\boldsymbol{f}_i}> P_\text{b}$, the algorithm proceeds by selecting the next branch $\boldsymbol{f}_{i+1}$ of the layer. 
After all possible valid branches in a given layer are evaluated, i.e., all valid $\boldsymbol{f}_i$ were fixed and its conditioned upper and lower bounds computed, the smallest known upper bound and its corresponding upper bound solution are updated as $\check{P}=\displaystyle\min_i (P_{\text{ub}|\boldsymbol{f}_i},\check{P})$ and $\check{\boldsymbol{\theta}}=\boldsymbol{\theta}_{\text{ub}|\boldsymbol{f}_i}$. With $\check{P}$ the algorithm proceeds to the pruning step where the set of approved branches, $\mathcal{G}_p^\prime$, in the current layer $p$ is constructed. The proposed pruning step aims to exclude from the search set all reflection coefficient vectors $\boldsymbol{\theta}$ that belong to $\chav{\boldsymbol{\theta}: 10\log_{10}\pc{\sfrac{P(\boldsymbol{\theta})}{P_\text{opt}}} > \gamma}$. This can implicitly be done by approving branches $\boldsymbol{f}_i$ such that $P_{\text{lb},\boldsymbol{f}_{i}} < (1-\delta) \check{P}$ with $\delta=1-10^{-\frac{\gamma}{10}}$. With this, the set of approved branches for the given layer $p$ is constructed as $\mathcal{G}_p^\prime=\{\boldsymbol{f}_{i}|P_{\text{lb},\boldsymbol{f}_{i}} < (1-\delta) \check{P}\}$. After pruning, the set of valid subvectors is updated and the algorithm repeats this process in the next layer. If the algorithm reaches the last layer, only a few valid candidate solutions are expected to remain. With this, they are all evaluated against $\check{P}$, and the optimal value is determined by the vector that yields the minimum value of $P$. 

\subsubsection{Subproblem Formulation}

A general formulation of the lower bounding subproblems of the proposed PBB algorithm is given in \eqref{opt:lb_subproblem_pre}. In what follows, the specific PSEP and PUBSEP subproblems are devised. Based on the PSEP problem from \eqref{opt:rv_slp_relaxed_sep} one rewrites \eqref{opt:lb_subproblem_pre} as 
 \begin{align}
\label{opt:lb_subproblem_sep_pre}
    \{P_{\text{lb}|\boldsymbol{f}_i}&,\boldsymbol{v}_{\text{r,lb}|\boldsymbol{f}_i}\}=\min_{\boldsymbol{v}_\text{r},P\in\mathbb{R}_+}  \ P \\
    &\hspace{-1.5em}\text{s.t.}\hspace{0.5em}  C(\boldsymbol{v}_\text{r}) \in \mathcal{J}, \ \ \notag\\
    &\hspace{-2.5em}-\displaystyle \sum_{\xi=1}^2 \text{ln}\pc{\Phi\pc{ \sqrt{\frac{P}{\sigma_w^2}}\pc{\boldsymbol{\kappa}_{\xi,k}^T \boldsymbol{f}_{\text{r},i}+\boldsymbol{\varrho}_{\xi,k}^T \boldsymbol{v}_{\text{r}} }}} \leq \beta_k, \ \ \text{for} \ k \in \mathcal{K}, \notag 
\end{align}
where $\mathcal{J}$ is the convex hull of $\mathcal{T}^{(N-p)}$, $\boldsymbol{\kappa}_{\xi,k}$ and $\boldsymbol{\varrho}_{\xi,k}$ correspond to the first $2p$, and, of the subsequent $2(N-p)$ entries of $\boldsymbol{h}_{\xi,k}$, respectively. The lower bounding PSEP subproblem is cast by rewriting \eqref{opt:lb_subproblem_sep_pre} using the auxiliary variable $\boldsymbol{\Tilde{x}}=[\sqrt{P}\boldsymbol{v}_\text{r}^T,\sqrt{P}]^T$, which yields 
\begin{align}
\label{opt:lb_subproblem_sep}
    \tilde{\boldsymbol{x}}_{\boldsymbol{f}_i}&=\min_{\tilde{\boldsymbol{x}}}  \ \boldsymbol{d}^T\Tilde{\boldsymbol{x}} \\
    &\hspace{-1.1em}\text{s.t.}\hspace{0.5em}   \pc{\boldsymbol{R}^\prime+\boldsymbol{R}_\text{fixed}\boldsymbol{f}_{\text{r},i}\boldsymbol{d}^T}\tilde{\boldsymbol{x}}\preceq \boldsymbol{0},\notag\\
    &\hspace{-1.5em} \displaystyle -\sum_{\xi=1}^2 \text{ln}\pc{ \Phi\pc{\frac{\boldsymbol{\kappa}_{\xi,k}^T \boldsymbol{f}_{\text{r},i}\boldsymbol{d}^T\tilde{\boldsymbol{x}}+\boldsymbol{\psi}_{\xi,k}^T \tilde{\boldsymbol{x}}}{\sigma_w}}} \leq \beta_k, \ \text{for} \ k\in \mathcal{K}, \notag  
\end{align}
where $\boldsymbol{d}=[\boldsymbol{0}_{2(N-p)}^T,1]^T$, $\boldsymbol{R}_\text{fixed}$ is composed of the first $2p$ columns of $\boldsymbol{R}$, $\boldsymbol{R}^\prime$ consists of the last $2(N-p)+1$ columns of $\boldsymbol{R}$ and $\boldsymbol{\psi}_{\xi,k}=[\boldsymbol{\varrho}_{\xi,k}^T,0]^T$. Similar steps can be applied with the PUBSEP formulation such that the PUBSEP lower bounding subproblem is written as
\begin{align}
\label{opt:lb_subproblem_ubsep}
    \tilde{\boldsymbol{x}}_{\boldsymbol{f}_i}&=\min_{\tilde{\boldsymbol{x}}}  \ \boldsymbol{d}^T\Tilde{\boldsymbol{x}} \\
    &\hspace{-1.1em}\text{s.t.}\hspace{0.5em}   \pc{\boldsymbol{D}^\prime+\boldsymbol{D}_\text{fixed}\boldsymbol{f}_{\text{r},i}\boldsymbol{d}^T}\tilde{\boldsymbol{x}}\preceq \boldsymbol{0},\notag\\
    &\hspace{-1.1em} \displaystyle \sum_{\xi=1}^2 \frac{1}{2}\text{erfc}\pc{ \frac{\boldsymbol{\eta}_{\xi,k}^T \boldsymbol{f}_{\text{r},i}\boldsymbol{d}^T\tilde{\boldsymbol{x}}+\boldsymbol{\zeta}_{\xi,k}^T \tilde{\boldsymbol{x}}}{\sigma_w}} \leq \rho_k, \ \text{for} \ k\in \mathcal{K}, \notag
\end{align}
where  $\boldsymbol{\zeta}_{\xi,k}=[\boldsymbol{\lambda}_{\xi,k}^T,0]^T$, $\boldsymbol{\eta}_{\xi,k}$ and $\boldsymbol{\lambda}_{\xi,k}$ correspond to the first $2p$ and of the subsequent $2(N-p)$ entries of $\boldsymbol{u}_{\xi,k}$, respectively, and,  $\boldsymbol{D}_\text{fixed}$ and $\boldsymbol{D}^\prime$ are composed of the first $2p$ and of the last $2(N-p)+1$ columns of $\boldsymbol{D}$, respectively. Solving \eqref{opt:lb_subproblem_sep} and \eqref{opt:lb_subproblem_ubsep} yields $\tilde{\boldsymbol{x}}_{\boldsymbol{f}_i}$ from which $\boldsymbol{v}_{\text{r,lb}|\boldsymbol{f}_i}$ and $P_{\text{lb}|\boldsymbol{f}_i}$ are readily extracted. The steps of the proposed PBB method are detailed in Algorithm \ref{alg:bnb}. \looseness -1
\begin{algorithm}[!ht]
\footnotesize
  \caption{Proposed Partial Branch-and-Bound Algorithm}
	\label{alg:bnb}
  \begin{algorithmic}    
  \State{\textbf{Inputs}: $\boldsymbol{h}_k$ for $k\in\mathcal{K}$, $\boldsymbol{s}$, $\mathcal{T}$, $P_\text{b}$, $\delta$, \textit{Criterion} \hspace{1em} \textbf{Output}: $\boldsymbol{\theta}_\text{out}$, $P_\text{out}$ }
  \vspace{1mm}
    \State{\textbf{If} \textit{Criterion} = PSEP: Solve \eqref{opt:power_min_sep} to get $\boldsymbol{x}_\text{lb}=[\sqrt{P_\text{lb}}\boldsymbol{\theta}_\text{r,lb}, \sqrt{P_\text{lb}}]^T$}
    \State{\textbf{If} \textit{Criterion} = PUBSEP: Solve \eqref{opt:power_min_ubsep} to get $\boldsymbol{x}_\text{lb}=[\sqrt{P_\text{lb}}\boldsymbol{\theta}_\text{r,lb}, \sqrt{P_\text{lb}}]^T$}
    \State{\textbf{If} $\boldsymbol{\theta}_\text{lb}=C(\boldsymbol{\theta}_\text{r,lb})\in \mathcal{T}^N$: \textbf{terminate with} $\boldsymbol{\theta}_\text{out}=\boldsymbol{\theta}_\text{lb}$ and $P_\text{out}=P_\text{lb}$ }
    \State{Compute $\boldsymbol{\theta}_\text{ub}=Q(\boldsymbol{\theta}_\text{lb})$ and get $\boldsymbol{\theta}_\text{r,ub}=R(\boldsymbol{\theta}_\text{ub})$}
    \State{\textbf{If} \textit{Criterion} = PSEP $\wedge \ \boldsymbol{h}_{\xi,k}^T\boldsymbol{\theta}_\text{r,ub}\geq0 \ \forall \ \xi\in\{1,2\}$ and $k\in\mathcal{K}$: Solve \eqref{opt:ub_power_sep} to get $P_\text{ub}$}
    \State{\textbf{Else If} \textit{Criterion} = PUBSEP $\wedge \ \boldsymbol{u}_{\xi,k}^T\boldsymbol{\theta}_\text{r,ub}\geq0 \ \forall \ \xi\in\{1,2\}$ and $k\in\mathcal{K}$: Solve \eqref{opt:ub_power_ubsep} to get $P_\text{ub}$}
    \State{\textbf{Else}: Set $\boldsymbol{\theta}_\text{ub}=\emptyset$ and $P_\text{ub}=\infty$} 
    \State{\textbf{If} $P_\text{ub}\leq P_\text{b}  \vee  10\log_{10}\pc{\sfrac{P_\text{ub}}{P_\text{lb}}}\leq \gamma$: \textbf{terminate with} $\boldsymbol{\theta}_\text{out}=\boldsymbol{\theta}_\text{ub}$ and $P_\text{out}=P_\text{ub}$}
    \vspace{1mm}
    \State{Define $\check{\boldsymbol{\theta}}=\boldsymbol{\theta}_\text{ub}$, $\check{P}=P_\text{ub}$ and the first level ($p=1$) of the tree by $\mathcal{G}_{p}:=\mathcal{T}_{\textrm{}}$}
	\For{$p=1:N-1$}
	\State{Partition  $\mathcal{G}_{p}$ in $\boldsymbol{f}_{1},\ldots,\boldsymbol{f}_{\PM{\mathcal{G}_{p}}}$ }  
	  \For{$i=1:\left| \mathcal{G}_{p} \right|$}
        \State{\textbf{If} \textit{Criterion} = PSEP }
		\State{\hspace{0.5em}Conditioned on $\boldsymbol{f}_{\text{r},i}=R(\boldsymbol{f}_{i})$ solve \eqref{opt:lb_subproblem_sep} to get $\tilde{\boldsymbol{x}}_{\boldsymbol{f}_i}$}
        \State{\hspace{0.5em}Extract $P_{\text{lb}|\boldsymbol{f}_i}$ and $\boldsymbol{v}_{\text{lb}|\boldsymbol{f}_i}=C(\boldsymbol{v}_{\text{r,lb}|\boldsymbol{f}_i})$}
        \State{\hspace{0.5em}Map $\boldsymbol{v}_{\text{ub}|\boldsymbol{f}_i}=Q(\boldsymbol{v}_{\text{lb}|\boldsymbol{f}_i})$ and get $\boldsymbol{\theta}_{\text{ub}|\boldsymbol{f}_i}=[\boldsymbol{f}_{i}^T,\boldsymbol{v}_{\text{ub}|\boldsymbol{f}_i}^T]^T$}
        \State{\hspace{0.5em}\textbf{If} $\boldsymbol{h}_{\xi,k}^T\boldsymbol{\theta}_{\text{ub}|\boldsymbol{f}_i}\geq0, \forall \xi\in\{1,2\}$ and $k\in\mathcal{K}$}
        \State{\hspace{0.75em} Based on $\boldsymbol{\theta}_{\text{r,ub}|\boldsymbol{f}_i}=R(\boldsymbol{\theta}_{\text{ub}|\boldsymbol{f}_i)}$ solve \eqref{opt:ub_power_sep} to get $P_{\text{ub}|\boldsymbol{f}_i}$}
        \State{\hspace{0.75em} \textbf{If} $P_{\text{ub}|\boldsymbol{f}_i}\leq P_\text{b}$:  \textbf{terminate with} $\boldsymbol{\theta}_\text{out}=\boldsymbol{\theta}_{\text{ub}|\boldsymbol{f}_i}$ and $P_\text{out}=P_{\text{ub}|\boldsymbol{f}_i}$}
        \State{\textbf{Else If} \textit{Criterion} = PUBSEP }
		\State{\hspace{0.5em}Conditioned on $\boldsymbol{f}_{\text{r},i}=R(\boldsymbol{f}_{i})$ solve \eqref{opt:lb_subproblem_ubsep} to get $\tilde{\boldsymbol{x}}_{\boldsymbol{f}_i}$}
        \State{\hspace{0.5em}Extract $P_{\text{lb}|\boldsymbol{f}_i}$ and $\boldsymbol{v}_{\text{lb}|\boldsymbol{f}_i}=C(\boldsymbol{v}_{\text{r,lb}|\boldsymbol{f}_i})$}
        \State{\hspace{0.5em}Map $\boldsymbol{v}_{\text{ub}|\boldsymbol{f}_i}=Q(\boldsymbol{v}_{\text{lb}|\boldsymbol{f}_i})$ and get $\boldsymbol{\theta}_{\text{ub}|\boldsymbol{f}_i}=[\boldsymbol{f}_{i}^T,\boldsymbol{v}_{\text{ub}|\boldsymbol{f}_i}^T]^T$}
        \State{\hspace{0.5em}\textbf{If} $\boldsymbol{u}_{\xi,k}^T\boldsymbol{\theta}_{\text{ub}|\boldsymbol{f}_i}\geq0, \forall \xi\in\{1,2\}$ and $k\in\mathcal{K}$}
        \State{\hspace{0.75em} Based on $\boldsymbol{\theta}_{\text{r,ub}|\boldsymbol{f}_i}=R(\boldsymbol{\theta}_{\text{ub}|\boldsymbol{f}_i)}$ solve \eqref{opt:ub_power_ubsep} to get $P_{\text{ub}|\boldsymbol{f}_i}$}
        \State{\hspace{0.75em} \textbf{If} $P_{\text{ub}|\boldsymbol{f}_i}\leq P_\text{b}$:  \textbf{terminate with} $\boldsymbol{\theta}_\text{out}=\boldsymbol{\theta}_{\text{ub}|\boldsymbol{f}_i}$ and $P_\text{out}=P_{\text{ub}|\boldsymbol{f}_i}$}
    \State{\textbf{end If}}
	\EndFor
    \State{Update $\check{P} =\min( \check{P}, P_{\text{ub}|\boldsymbol{f}_i})$ and update $\check{\boldsymbol{\theta}}$ accordingly} 
	\State{Based on $\check{P}$ and on the lower bounds build the set:}
    \State{\hspace{0.5em}$\mathcal{G}_{p}^{\prime}:=\{  \boldsymbol{f}_i\vert\ P_{\textrm{lb}|\boldsymbol{f}_i} < (1-\delta) \check{P}, \ i= 1,\ldots,  \left|\mathcal{G}_{p}\right|\} $}
	\State{Define the set for the next level in the tree: $\mathcal{G}_{p+1}:=\mathcal{G}_{p}^{\prime} \times \mathcal{T}_{\textrm{}}$}
	\EndFor
    \For{$i=1:\left| \mathcal{G}_{N} \right|$}
    \State{\textbf{If} \textit{Criterion} = PSEP $\wedge\ \boldsymbol{h}_{\xi,k}^T\boldsymbol{\theta}_i\geq0$:} 
    \State{\hspace{0.5em}Solve \eqref{opt:ub_power_sep} with $\boldsymbol{\theta}_\text{r,ub}=R(\boldsymbol{\theta}_i)$ and get $P_i$}
    \State{\textbf{If} \textit{Criterion} = PUBSEP $\wedge\ \boldsymbol{u}_{\xi,k}^T\boldsymbol{\theta}_i\geq0$:}
    \State{\hspace{0.5em}Solve \eqref{opt:ub_power_ubsep} with $\boldsymbol{\theta}_\text{r,ub}=R(\boldsymbol{\theta}_i)$ and get $P_i$}
    \State{\textbf{If} $P_i\leq P_\text{b}$: \textbf{terminate with} $\boldsymbol{\theta}_\text{out}=\boldsymbol{\theta}_i$ and $P_\text{out}=P_i$ }
    \EndFor
	\State{The transmit power reads as $P_\text{out}=\min_{i\in\{1,\hdots,\vert\mathcal{G}_N\vert\}}(\check{P},P_i)$ and the reflection coefficients are given by $\boldsymbol{\theta}_{\textrm{out}} =\boldsymbol{\theta}_i$ }
\end{algorithmic}
\end{algorithm}

\subsection{\textcolor{r2}{On the Computational Complexity of the Algorithm and Latency-Aware Implementation}}

The initialization problems, described in \eqref{opt:power_min_sep} and \eqref{opt:power_min_ubsep}, the upper bounding problems from \eqref{opt:ub_power_sep} and \eqref{opt:ub_power_ubsep}, and lower bounding subproblems, described in \eqref{opt:lb_subproblem_sep} and \eqref{opt:lb_subproblem_ubsep}, are convex with twice continuously differentiable real-valued functions. With this, according to \cite[Chapter 11]{Boyd_2004}, they are solvable with the barrier method. 
The upper bound complexity order (UBCO) of the barrier method can be summarized as $\mathcal{O}(\sqrt{\varphi}q^3)$, \cite{lopes_tcom2023},\cite[Section 11.5.6]{Boyd_2004}, with $\varphi$ being the number of inequality constraints and $q$ being the number of optimization variables.
By substituting the values of $\varphi$ and $q$ for the different problems one reaches the conclusion that solving \eqref{opt:power_min_sep} and \eqref{opt:power_min_ubsep} yields UBCOs of $\mathcal{O}(N^{3.5}+N^3\sqrt{K})$, \eqref{opt:ub_power_sep} and \eqref{opt:ub_power_ubsep} yields UBCOs of $\mathcal{O}(\sqrt{K})$, and \eqref{opt:lb_subproblem_sep} and \eqref{opt:lb_subproblem_ubsep} yields UBCOs of $\mathcal{O}(N^{3.5}+N^3\sqrt{K})$. For executing Algorithm \ref{alg:bnb}, it is necessary to solve the initialization problem once and the upper bounding problems and lower bounding subproblems, $J$ and $B$ times, respectively. Since by the design of the algorithm $J\leq \alpha_\theta B$, the UBCO of the proposed PBB algorithms is given by $\mathcal{O}(B (N^{3.5}+N^3\sqrt{K}))$.

\textcolor{r2}{As discussed in section \ref{sec:numerical_results_ris}, the value of ${B}$ remains small for systems with $P_\text{opt}\ll P_\text{b}$. Yet, one might argue that, in the worst-case scenario, the latency of PBB equals that of FBB. While PBB can indeed reach FBB complexity, applying the concept of parallelization to branch-and-bound methods reduces the latency added per layer to that of solving a single optimization problem. This holds because, as breadth-first search is used and each branch is computed independently, the optimization problems corresponding to each branch can be computed in parallel. Consequently, the latency per layer equals that of solving only one convex optimization problem (with a UBCO of $\mathcal{O}(N^{3.5}+N^3\sqrt{K})$), and the overall latency of branch-and-bound can be bounded by $N$ times this corresponding latency.}


\subsection{\textcolor{r1}{Operation Scenarios}}

\textcolor{r1}{When utilizing the proposed PBB approach $P_\text{b}$ and $\gamma$ are set based on the operation scenario. To compute the minimum transmit power required for attainment of the SEP requisites $P_\text{b} = 0$ and $\gamma = 0$ dB are set, representing the FBB performance, along with its associated computational complexity. 
However, there are scenarios where the computational demands of the FBB approach may be prohibitive. In such cases, performance-complexity trade-offs can be achieved by adjusting the values of $P_\text{b}$ and $\gamma$.}

\textcolor{r1}{
Setting $P_\text{b}$ involves accepting any suboptimal solution that satisfies $P \leq P_\text{b}$. In practical scenarios, well-designed systems operate with reasonable power margins to provide some robustness to the system, such that $P_\text{opt}\ll P_\text{max}$, where $P_\text{max}$ represents the maximum transmit power, e.g., present due to regulatory reasons. In this context, the value of $P_\text{b}$ is chosen as the maximum average transmit power that is considered adequate to spend (considering the corresponding costs). For this choice one should obey the condition $P_\text{opt}\ll P_\text{b}$, as for this case, it is expected that suboptimal solution pairs $(P_\text{out},\boldsymbol{\theta}_\text{out})$ that attain the users' SEP requisites can be found with few branch explorations which contributes for computational complexity reduction. On the other hand, the maximum value recommended for the target power budget is given by $P_\text{b}=P_\text{max}$ as, in most cases, one cannot surpass the $P_\text{max}$ and, thus, higher computational costs are required for computing $(P_\text{out},\boldsymbol{\theta}_\text{out})$ such that $P_\text{out}\leq P_\text{max}$.}

\textcolor{r1}{The determination of $\gamma$ reflects the acceptable tolerance for an increase in transmit power relative to $P_{\text{opt}}$. For regular operation with $\overline{P}_\text{opt}\ll P_\text{b}$, the value of $\gamma$ can be thought of as a soft margin of the system. In this sense, setting $\gamma=0$ would mean, for the case of $P_{\text{opt}}=P_\text{b}$, that $P_\text{b}$ is a hard limit and FBB complexity must be utilized for maintaining the target power budget of the system. On the other hand, setting $\gamma>0$ allows for soft power budgets and computation of complexity performance trade-offs.
}

\section{\textcolor{r1}{Problem Formulation and Algorithm Design for High-Resolution RIS}}
\label{sec:hr_ris}

\textcolor{r1}{To derive a low-complexity approach for the case of high-resolution reflecting elements this section considers that, for a sufficiently large $\alpha_\theta$, the discrete set $\mathcal{T}$ can be well approximated by $\mathcal{C}=\{\theta: \PM{\theta}^2=1\}$. 
With this, the PSEP and PUBSEP problems are reformulated based on the approximation of the original discrete set $\mathcal{T}$ by $\mathcal{C}$. Substituting $\mathcal{T}$ by $\mathcal{C}$ for the PSEP problem yields the following optimization problem}
\begin{align}
\label{opt:pmhr_sep_complex}
    &\min_{\boldsymbol{\theta}, P\in \mathbb{R}_+}  \ P \\
    &\hspace{0.5em}\text{s.t.} \ \PM{\pr{\boldsymbol{\theta}}_n}^2=1,\ \text{for} \ n \in \mathcal{N},\notag\\
&\hspace{2.5em}-\sum_{\xi=1}^2\text{ln}\pc{\Phi\pc{\sqrt{\frac{P}{\sigma_w^2}}{v}_{\xi,k}(\boldsymbol{\theta})}} \leq \beta_k, \ \text{for} \ k\in \mathcal{K}. \notag
\end{align}
The optimization problem of power minimization for high-resolution RIS under SEP constraints (PHSEP) is cast by rewriting \eqref{opt:pmhr_sep_complex} with real-valued variables which yield 
\begin{align}
\label{opt:pmhr_sep}
    &\min_{\boldsymbol{\Theta}\in \mathcal{M},P\in \mathbb{R}_+}  \ P \\
   &\hspace{0.5em} \displaystyle -\sum_{\xi=1}^2 \text{ln}\pc{\Phi\pc{\sqrt{\frac{P}{\sigma_w^2}}\ \text{tr}\pc{\boldsymbol{\Theta}  \boldsymbol{H}_{\xi,k} }} } \leq \beta_k, \ \text{for} \ k\in \mathcal{K}, \notag
\end{align}
\textcolor{r1}{where $\mathcal{M}=\{\boldsymbol{\Theta} \in \mathbb{R}^{2\times N}:  [\boldsymbol{\Theta}^T\boldsymbol{\Theta}]_{(n,n)}=1, \ \text{for} \ n\in \mathcal{N}\}$,
\begin{align}
\boldsymbol{H}_{1,k}&=\begin{bmatrix}
    \text{Re}\{\boldsymbol{b}^T_{\text{R},k}\} \\
  - \text{Im}\{\boldsymbol{b}^T_{\text{R},k}\} 
\end{bmatrix}^T, 
\quad
\boldsymbol{H}_{2,k}=\begin{bmatrix}
    \text{Im}\{\boldsymbol{b}_{\text{I},k}^T\}\\
    \text{Re}\{\boldsymbol{b}_{\text{I},k}^T\}
\end{bmatrix}^T, \\ 
\boldsymbol{\Theta}&=\begin{bmatrix}
    \pc{\text{Re}\{\boldsymbol{\theta}\}}^T \\
    \pc{\text{Im}\{\boldsymbol{\theta}\}}^T
\end{bmatrix}, 
\quad \boldsymbol{b}_{\text{R},k}=r_k\boldsymbol{h}_k^H, \quad \boldsymbol{b}_{\text{I},k}=i_k\boldsymbol{h}_k^H. 
\end{align}
}
Similarly, substituting $\mathcal{T}$ by $\mathcal{C}$ with PUBSEP formulation yields the following problem
\begin{align}
\label{opt:complex_form}
    &\min_{\boldsymbol{\theta}, P}  \ P \\
    &\hspace{0.7em}\text{s.t.}\hspace{0.5em} \PM{\pr{\boldsymbol{\theta}}_n}^2=1,\ \text{for} \ n \in \mathcal{N}, \hspace{1em} P\geq 0, \notag \\
    &\hspace{0.7em} \sum_{\xi=1}^2 \frac{1}{2}\text{erfc}\pc{\frac{d_{\xi,k}\pc{\boldsymbol{\theta},P}}{\sigma_w}} \leq \rho_k, \ \text{for} \ k\in \mathcal{K}.\notag 
\end{align}
The problem for power minimization for high-resolution RIS under UBSEP constraints (PHUBSEP) is cast by rewriting \eqref{opt:complex_form} with real-valued variables which yields 
\begin{align}
\label{opt:rv_slp_unconstrained}
    &\min_{\boldsymbol{\Theta}\in \mathcal{M},P\in \mathbb{R}_+}  \ P \\
   &\hspace{0.5em} \displaystyle \sum_{\xi=1}^2 \frac{1}{2}\text{erfc}\pc{ {\sqrt{\frac{P}{\sigma_w^2}}}\ \text{tr}\pc{\boldsymbol{\Theta}  \boldsymbol{U}_{\xi,k} } } \leq \rho_k, \ \text{for} \ k\in \mathcal{K}. \notag
\end{align}
where $\boldsymbol{a}_k=s_k^*\boldsymbol{h}_k^H$, and, 
\begin{align}
    \boldsymbol{U}_{1,k}&=\begin{bmatrix}
        \text{Re}\{\boldsymbol{a}^T_k\} \sin(\phi) - \text{Im}\{\boldsymbol{a}^T_k\} \cos(\phi)\\
        -\text{Re}\{\boldsymbol{a}^T_k\} \cos(\phi) - \text{Im}\{\boldsymbol{a}^T_k\} \sin(\phi)
    \end{bmatrix}^T, \notag\\
    \boldsymbol{U}_{2,k}&=\begin{bmatrix}
        \text{Re}\{\boldsymbol{a}_k^T\} \sin(\phi) + \text{Im}\{\boldsymbol{a}_k^T\} \cos(\phi)\\
        \text{Re}\{\boldsymbol{a}_k^T\} \cos(\phi) - \text{Im}\{\boldsymbol{a}_k^T\} \sin(\phi)
    \end{bmatrix}^T. \notag
\end{align}
As demonstrated in Appendix \ref{app:high_res}, the SEP constraint functions are convex and, although the UBSEP functions are not geodesically convex in $\mathcal{M}$, one can restrict the feasible set such that UBSEP functions are geodesically convex. Yet, due to the set $\mathcal{M}$ not being geodesically convex \cite[section 2.3]{boumal2023intromanifolds}, the optimization problems from \eqref{opt:pmhr_sep} and \eqref{opt:rv_slp_unconstrained} are not geodesically convex, implying that the application of descent methods only guarantees local optimality.

\subsection{Local Optimum via the Proposed Bisection Method}
\label{sec:bisection}
A locally optimal solution for the proposed high-resolution problems is computed via the proposed bisection method. The method is initialized with $P_-$ as a lower bound on $P_\text{lopt}$ and $P_+$ as an upper bound on $P_\text{lopt}$. The variable $P$ is fixed as $P_0=\sfrac{(P_+ + P_-)}{2}$ and the remaining problem's feasibility is evaluated. If feasible, $P_+$ is updated as $P_0$, otherwise, $P_-$ is updated as $P_0$. This is done recursively until the power difference between two consecutive iterations is below an optimality tolerance $\epsilon_\text{tol}$. For a given $P$, the general optimization problem is written as 
\begin{align}
\label{opt:sub}
    &\displaystyle \text{find}_{\boldsymbol{\Theta} \in \mathcal{M}}\ \ \boldsymbol{\Theta} \\
    &\hspace{0.5em}\text{s.t.} \  f_k(\boldsymbol{\Theta})\leq 0, \ \text{for} \ k\in \mathcal{K}, \notag
\end{align}
where, for the PHSEP formulation $
    f_k(\boldsymbol{\Theta})=-\sum_{\xi=1}^2 \text{ln}\pc{\Phi\pc{\sqrt{\frac{P}{\sigma_w^2}}\ \text{tr}\pc{\boldsymbol{\Theta}  \boldsymbol{H}_{\xi,k} }} }-\beta_k,$
and for the PHUBSEP formulation $
    f_k(\boldsymbol{\Theta})=\sum_{\xi=1}^2 \frac{1}{2}\text{erfc}\pc{ {\sqrt{\frac{P}{\sigma_w^2}}}\ \text{tr}\pc{\boldsymbol{\Theta}  \boldsymbol{U}_{\xi,k} } } -\rho_k.$
The strategy for solving \eqref{opt:sub} consists of minimizing the maximum constraint function and evaluating if the locally optimal solution attains it. This yields the following problem
\begin{align}
\label{opt:subproblem2}
    \boldsymbol{\Theta}_\text{lopt}&=\displaystyle \min_{\boldsymbol{\Theta \in \mathcal{M}}}\ \max_{k\in \mathcal{K}} f_k(\boldsymbol{\Theta}). 
\end{align}
Based on $\boldsymbol{\Theta}_\text{lopt}$ the feasibility of \eqref{opt:subproblem2} is evaluated by checking $f(\boldsymbol{\Theta}_\text{lopt})\leq 0$, where
\begin{align}
\label{eq:f}
     f(\boldsymbol{\Theta})=\max_{k\in \mathcal{K}} \ f_k(\boldsymbol{\Theta}) .
\end{align}   
If the condition holds the problem is feasible and $\boldsymbol{\Theta}_\text{lopt}$ is a solution of \eqref{opt:sub}. Otherwise, at least one constraint cannot be fulfilled with the given transmit power $P$, implying that \eqref{opt:sub} is infeasible. The steps of the BM are further detailed in algorithm \ref{alg:bisection}.

\begin{algorithm}[t]
\small
  \caption{Proposed Bisection Method}
	\label{alg:bisection}
  \begin{algorithmic}    
  \State{\textbf{Inputs}: ${P}_+ \geq P_\text{opt}$, ${P}_-\leq P_\text{opt}$, $f_a<0$, $i_\text{max}>0$, $\epsilon_\text{tol}>0$ \hspace{1em} \textbf{Output}: $P_\text{opt}$, $\boldsymbol{\theta}_\text{opt}$ }
  \State{Define $i=0$, $P_a=P_0$}
  \State{\textbf{While} $\pc{ {P_{a}-P_{0}} \leq \epsilon_\text{tol}\ \vee \ i\leq i_\text{max}} \ \wedge \ f_a< 0$}
  \State{\hspace{0.5em} Solve \eqref{opt:subproblem3} with RCG \cite{boumal_manopt} considering $P=P_0$ and get $\boldsymbol{\Theta}_\text{opt}$}
  \State{\hspace{0.5em} Compute $f_a=f(\boldsymbol{\Theta}_\text{opt})$ with \eqref{eq:f}}
  \State{\hspace{0.5em} \textbf{If} $f_a\leq0$ $\rightarrow$ Update $P_{+}={P_0}$}
  \State{\hspace{0.5em} \textbf{Else} $\rightarrow$ Update $P_{-}={P_0}$}
 \State{\hspace{0.5em} Update $P_a=P_0$, $P_0=\sfrac{\pc{P_+ + P_-}}{2}$ and $i={i+1}$}
\State{ Update $P_\text{opt}=P_0$ and $\boldsymbol{\theta}_\text{opt}=\pr{\boldsymbol{\Theta}_\text{opt}}_{(1,:)}+j\pr{\boldsymbol{\Theta}_\text{opt}}_{(2,:)}$}
    \end{algorithmic}
\end{algorithm}

\subsubsection{Evaluating Feasibility via Riemannian Conjugate Gradient}

The utilization of algorithm \ref{alg:bisection} implies a method for locally solving the unconstrained problem in \eqref{opt:subproblem2}. This is done with the RCG algorithm \cite[Section 3.1]{boumal2014optimization}, designed for solving unconstrained minimization problems in Riemannian manifolds.
Since the RCG approach requires a twice continuously differentiable objective, $f(\boldsymbol{\Theta})$ is substituted by its softmax approximation computed with the log-sum-exp function $\text{LSE}(\boldsymbol{x})=\text{ln}\pc{\sum_{i} e^{x_i}}$, which yields $
    f_0 (\boldsymbol{\Theta})=\text{ln}\pc{\sum_{k=1}^K e^{f_k(\boldsymbol{\Theta})}}.$
Note that, $\text{LSE}(\boldsymbol{x})$ is a non-decreasing function, which implies that for the regions where $f(\boldsymbol{\Theta})$ is convex the convexity is preserved. With this, \eqref{opt:subproblem2} is rewritten as
\begin{align}
\label{opt:subproblem3}
    & \min_{\boldsymbol{\Theta \in \mathcal{M}}}\ \text{ln}\pc{ \sum_{k=1}^K e^{f_k(\boldsymbol{\Theta})}}.
\end{align}
A locally optimal solution to the optimization problem described in \eqref{opt:subproblem3} is computed via the RCG algorithm. To this end, however, the Euclidean gradient of $f_0(\boldsymbol{\Theta})$ and a strictly feasible starting point $\boldsymbol{\Theta}_0$ are required. In what follows these values are computed for both design criteria.

\paragraph{PHSEP RCG}
With the PHSEP formulation the Euclidean gradient $\nabla f_0(\boldsymbol{\Theta})$ is given by
\begin{align}
    &\nabla f_0(\boldsymbol{\Theta})=\pc{\sum_{k=1}^K{e^{f_k(\boldsymbol{\Theta})}\nabla f_k}}\pc{\sum_{k=1}^K e^{f_k(\boldsymbol{\Theta})}}^{-1}, \notag\\
    &\nabla f_k(\boldsymbol{\Theta})= -\sqrt{\frac{P}{2\pi\sigma_w^2}}\sum_{\xi=1}^2 \frac {e^{- \frac{P}{2\sigma_w^2}{\text{tr}(\boldsymbol{\Theta}  \boldsymbol{H}_{\xi,k})^2} } \boldsymbol{H}_{\xi,k}^T}{\Phi\pc{\sqrt{\frac{P}{\sigma_w^2}}\ \text{tr}\pc{\boldsymbol{\Theta}  \boldsymbol{H}_{\xi,k}}}}. \notag
\end{align}
\textcolor{r2}{As demonstrated in Appendix \ref{app:high_res}, the SEP functions $f_k(\boldsymbol{\Theta})$ are convex in $\mathbb{R}^{2\times N}$, which implies that $f_0(\boldsymbol{\Theta})$ is also convex.} With this, initializing the RCG algorithm with any $\boldsymbol{\Theta}_0 \in \mathcal{M}$ guarantees convergence to a locally optimal solution.

\paragraph{PHUBSEP RCG}
For PHUBSEP the values of the Euclidean gradient read as 
\begin{align}
    &\nabla f_0(\boldsymbol{\Theta})=\pc{\sum_{k=1}^K{e^{f_k(\boldsymbol{\Theta})}\nabla f_k}}\pc{\sum_{k=1}^K e^{f_k(\boldsymbol{\Theta})}}^{-1}, \notag\\
    &\nabla f_k(\boldsymbol{\Theta})=- \sqrt{\frac{P}{\pi\sigma_w^2}}\sum_{\xi=1}^2 e^{- {\frac{P}{\sigma_w^2}\cdot \text{tr}\pc{\boldsymbol{\Theta}  \boldsymbol{U}_{\xi,k} }^2} } \boldsymbol{U}_{\xi,k}^T. \notag
\end{align}
\textcolor{r2}{As demonstrated in Appendix \ref{app:high_res}, the functions $f_k(\boldsymbol{\Theta})$ are convex for $\boldsymbol{\Theta}\in \mathcal{Y}$ with $\mathcal{Y}=\{\boldsymbol{\Theta}: \text{tr}\pc{\boldsymbol{\Theta} \boldsymbol{U}_{\xi,k}}\geq0,\  \text{for}\ k \in \mathcal{K}, \ \xi \in\{1,2\}\}$, 
which implies convexity of $f_0(\boldsymbol{\Theta})$ for $\boldsymbol{\Theta}\in \mathcal{Y}$.}
Due to $f_0(\boldsymbol{\Theta})$ being convex for $\boldsymbol{\Theta}\in \mathcal{Y}$, if the optimal solution of \eqref{opt:subproblem3}, termed $\boldsymbol{\Theta}_\text{lopt}$, belongs to $\mathcal{Y}$, initializing the RCG method with any value of $\boldsymbol{\Theta}_0 \in\mathcal{Y}$ supports finding a locally optimal solution. This is the case since $f_0(\boldsymbol{\Theta})$ grows for a decrease in $\text{tr}\pc{\boldsymbol{\Theta} \boldsymbol{U}_{\xi,k}}$ and thus by initializing the RCG algorithm with $\boldsymbol{\Theta}_0 \in\mathcal{Y}$ it will takes steps to stay on $\mathcal{Y}$.
On the other hand, if $\boldsymbol{\Theta}_\text{lopt}\notin \mathcal{Y}$, which corresponds to a noiseless received signal outside the correct decision region, the value of at least one constraint function $f_k(\boldsymbol{\Theta})$ is given by $f_k(\boldsymbol{\Theta})\geq 0.5-\rho_k$. Since $f(\boldsymbol{\Theta})=\max_{k} f_k(\boldsymbol{\Theta})$, this implies that in this case \eqref{opt:rv_slp_unconstrained} is infeasible for $\rho_k<0.5 \ \forall k\in\mathcal{K}$. For this case, initializing the RCG algorithm for solving \eqref{opt:subproblem3} with any starting point, including $\boldsymbol{\Theta}_0 \in\mathcal{Y}$, will yield an output $\boldsymbol{\Theta}_\text{out}$ such that $f(\boldsymbol{\Theta}_\text{out})>0$.
With this, the optimization problem for computing the initial point $\boldsymbol{\Theta}_0\in \mathcal{Y}$ can be cast as 
\begin{align}
\label{opt:trace}
    &\displaystyle \max_{\boldsymbol{\Theta \in \mathcal{M}}}\ \min_{k \in \mathcal{K}, \xi \in \{1,2\}} \text{tr} \pc{\boldsymbol{\Theta}  \boldsymbol{U}_{\xi,k} }.
\end{align}
To solve \eqref{opt:trace} via the RCG algorithm the log-sum-exp function is applied which yields
\begin{align}
\label{opt:initial_problem}
    &\displaystyle \boldsymbol{\Theta}_0=\min_{\boldsymbol{\Theta \in \mathcal{M}}}\ v_0(\boldsymbol{\Theta}),
\end{align}
with $v_0(\boldsymbol{\Theta})=\text{ln}\pc{ \sum_{k=1}^K e^{-\text{tr}\pc{\boldsymbol{\Theta}  \boldsymbol{U}_{1,k} }} + e^{-\text{tr}\pc{\boldsymbol{\Theta}  \boldsymbol{U}_{2,k} }}}$, and, $
    \nabla v_0(\boldsymbol{\Theta})=- \frac{\sum_{k=1}^K e^{-\text{tr}\pc{\boldsymbol{\Theta}  \boldsymbol{U}_{1,k} }}\boldsymbol{U}_{1,k}^T+e^{-\text{tr}\pc{\boldsymbol{\Theta}  \boldsymbol{U}_{2,k} }}\boldsymbol{U}_{2,k}^T}
    {\sum_{k=1}^K e^{-\text{tr}\pc{\boldsymbol{\Theta}  \boldsymbol{U}_{1,k} }} + e^{-\text{tr}\pc{\boldsymbol{\Theta}  \boldsymbol{U}_{2,k} }}}.
$
Problem \eqref{opt:initial_problem} is solved via the utilization of the RCG algorithm with any starting point $\boldsymbol{\Theta}\in \mathcal{M}$. The details of the RCG implementation are given in \cite{boumal_manopt}.

\subsubsection{Final Considerations}

The complexity of the RCG algorithm dominates the complexity of the proposed PHSEP and PHUBSEP methods. Considering that the computational cost of computing the Euclidean gradient is $\mathcal{O}\pc{N^2}$ and that the number of iterations required for convergence scales with $\sqrt{N}$, one can summarize the UBCO of the RCG algorithm as $\mathcal{O}\pc{N^{2.5}}$.
Since the number of times that RCG is required to run mainly depends on initialization of $P_+$ and $P_-$ and does not grow with the size of the system the overall UBCO of the proposed algorithm is in the order of $\mathcal{O}\pc{N^{2.5}}$.

The approximation of $\mathcal{T}$ by $\mathcal{C}$ implies that the RIS has sufficiently high resolution such that formulating the problem with a continuous set is beneficial for achieving a reasonable solution. Yet, in practice, the reflection coefficients are constraints to a discrete set since although large the resolution is always finite. By applying algorithm \ref{alg:bisection}, one computes $P_\text{lopt}\in \mathbb{R}_+$ and $\boldsymbol{\theta}_{\text{lopt}}\in \mathcal{C}^N$. Note that, since $\boldsymbol{\theta}_{\text{lopt}}$ does not necessarily belong to $\mathcal{T}^{N}$ a projection step is necessary. This is done, similarly as in section \ref{sec:precoding_design}, via UQ such that $\boldsymbol{\theta}=Q(\boldsymbol{\theta}_{\text{lopt}})$. Accordingly, the transmit power $P$ is computed by solving \eqref{opt:ub_power} with the corresponding formulation and $\boldsymbol{\theta}_\text{r,ub}=R(\boldsymbol{\theta})$. After the mentioned steps $\boldsymbol{\theta}\in \mathcal{T}^N$ and $P\in\mathbb{R}_+$ can be utilized for transmission.

\section{\textcolor{r1}{Formulation of Worst-Case Optimization for Imperfect CSI}}
\label{sec:csi}

\textcolor{r1}{Given the availability of channel estimation approaches, e.g., \cite{Araújo_cest_jstsp,Zhang_cest_jsac}, perfect CSI at the RIS controller was considered in previous sections.}
\textcolor{r1}{This section expands the proposed formulations to imperfect CSI scenarios while maintaining compatibility with the algorithms developed in sections \ref{sec:precoding_design} and \ref{sec:bisection}. This is done by addressing worst-case optimization problems for a given CSI imperfection. To this end, the ellipsoid channel mismatch model  \textcolor{r2}{\cite{zheng2008robust,Schober_jsac2020,zhou2022cooperative,ellipsoid_Jiaheng}} is assumed. With this, the channels $\boldsymbol{h}_k$ for $k \in \mathcal{K}$ are modeled as $\boldsymbol{h}_k=\hat{\boldsymbol{h}}_k+\tilde{\boldsymbol{h}}_k$, where $\hat{\boldsymbol{h}}_k$ is the estimated channel and $\tilde{\boldsymbol{h}}_k$ is the unknown part of the channel model. For formulation of the worst-case $\tilde{\boldsymbol{h}}_k\sim\mathcal{CN}(0,\sigma_{h,k}^2 \boldsymbol{I})$ is assumed to belong to $\Omega_k$, with $\Omega_k= \chav{\boldsymbol{v} \in \mathbb{C}^{M \times 1} : \|\boldsymbol{v}\|_2 \leq \epsilon_k}$.} 

\subsection{\textcolor{r1}{Derivation of the SEP-based Worst-Case Optimization}}

Applying the channel model to the $k$-th user's SEP function yields $ \text{P}\pc{\hat{s}_k\neq s_k|\boldsymbol{\theta},P}=-\sum_{\xi=1}^2\ln\pc{\Phi\pc{ \pc{\sqrt{\sfrac{P}{\sigma_w^2}}} \pc{\hat{v}_{\xi,k}\pc{\boldsymbol{\theta}}+\tilde{v}_{\xi,k}\pc{\boldsymbol{\theta}}}}}$, with $\hat{v}_{1,k}\pc{\boldsymbol{\theta}}= r_k \text{Re}\{\hat{\boldsymbol{h}}_k^H \boldsymbol{\theta}\}$, $\hat{v}_{2,k}(\boldsymbol{\theta})=i_k \text{Im}\{\hat{\boldsymbol{h}}_k^H \boldsymbol{\theta}\}$, $\tilde{v}_{1,k}\pc{\boldsymbol{\theta}}=r_k \text{Re}\{\tilde{\boldsymbol{h}}_k^H \boldsymbol{\theta}\}$, 
$\tilde{v}_{2,k}(\boldsymbol{\theta})= i_k \text{Im}\{\tilde{\boldsymbol{h}}_k^H \boldsymbol{\theta}\}$.
For given $\boldsymbol{\theta}$ and $P$, the vector $\tilde{\boldsymbol{h}}_{k}$ which yields the $k$-th user largest SEP (LSEP) is determined by solving the following problem
\begin{align}
\label{opt:lsep_ncov}
    &\text{LSEP}(\tilde{\boldsymbol{h}}_{k})=\max_{\tilde{\boldsymbol{h}}_{k}}  \ -\sum_{\xi=1}^2\ln\pc{\Phi\pc{\sqrt{\frac{P}{\sigma_w^2}}\pc{ \hat{v}_{\xi,k}+\tilde{v}_{\xi,k}(\tilde{\boldsymbol{h}}_k)}}}\notag\\
    &\hspace{0.5em}\text{s.t.}\hspace{0.5em}\tilde{v}_{1,k}(\tilde{\boldsymbol{h}}_k)= r_k \text{Re}\{\tilde{\boldsymbol{h}}_k^H \boldsymbol{\theta}\}, \hspace{0.5em} 
    \tilde{v}_{2,k}(\tilde{\boldsymbol{h}}_k)= i_k \text{Im}\{\tilde{\boldsymbol{h}}_k^H \boldsymbol{\theta}\}, \\
    &\hspace{2.5em}\|\tilde{\boldsymbol{h}}_{k}\|_2\leq {\epsilon_k}. \notag
\end{align}
Problem \eqref{opt:lsep_ncov} is the maximization of a convex function, which corresponds to a hard-to-solve non-convex problem. 
 
\begin{align}
    f_k(\boldsymbol{\theta})
    &=\displaystyle -\sum_{\xi=1}^2\ln\pc{\Phi\pc{\sqrt{\frac{P}{\sigma_w^2}}\pc{ \hat{v}_{\xi,k}\pc{\boldsymbol{\theta}}-\sqrt{2N}\epsilon_k}} }.
\end{align}
Based on $f_k(\boldsymbol{\theta})$ the power minimization problem can be rewritten as 
\begin{align}
\label{opt:cv_psep_csi}
    &\min_{\boldsymbol{\theta} \in \mathcal {T}^N,P\in \mathbb{R}_+}  \ P \\
    &\text{s.t.} -\hspace{-0.2em}\sum_{\xi=1}^2\ln\pc{\Phi\pc{\sqrt{\frac{P}{\sigma_w^2}}\pc{ \hat{v}_{\xi,k}\pc{\boldsymbol{\theta}}-\sqrt{2N}\epsilon_k}}}\hspace{-0.2em}\leq \beta_k,  \text{for} \ k\in \mathcal{K}.
    \notag
\end{align}
By rewriting problem \eqref{opt:cv_psep_csi} with real-valued variables, one constructs the proposed PSEP problem for imperfect CSI as 
\begin{align}
    &\min_{\boldsymbol{\theta}_\text{r},P\in \mathbb{R}_+}  \ P \\
    &\hspace{0.5em}\text{s.t.}\hspace{0.5em} \pr{\boldsymbol{\theta}_\text{r}}_{2n-1}+ j\pr{\boldsymbol{\theta}_\text{r}}_{2n} \in \mathcal{T}, \ \text{for} \ n\in \mathcal{N}, \ \  \notag \\
    &\hspace{0.5em} \displaystyle -\sum_{\xi=1}^2 \text{ln}\pc{ \Phi\pc{\sqrt{\frac{P}{\sigma_w^2}}\pc{\hat{\boldsymbol{h}}_{\xi,k}^T \boldsymbol{\theta}_{\text{r}}- \sqrt{2N}\epsilon_k}} }\leq \beta_k, \ \text{for} \ k\in \mathcal{K}, \notag
\end{align}
with $\hat{\boldsymbol{h}}_{\text{1},k}^T=\pr{[\hat{\boldsymbol{\gamma}}_{R,k}]_1, -[\hat{\boldsymbol{\gamma}}_{I,k}]_1,\hdots, [\hat{\boldsymbol{\gamma}}_{R,k}]_N, -[\hat{\boldsymbol{\gamma}}_{I,k}]_N }$, $\hat{\boldsymbol{h}}_{\text{2},k}^T=\pr{[\hat{\boldsymbol{\gamma}}_{I,k}]_1, [\hat{\boldsymbol{\gamma}}_{R,k}]_1,\hdots, [\hat{\boldsymbol{\gamma}}_{I,k}]_N, [\hat{\boldsymbol{\gamma}}_{R,k}]_N}$, $\hat{\boldsymbol{\gamma}}_{R,k}=r_k\text{Re}\{\hat{\boldsymbol{h}}_k^H\}$, $\hat{\boldsymbol{\gamma}}_{I,k}=i_k\text{Im}\{\hat{\boldsymbol{h}}_k^H\}$.
Departing from \eqref{opt:cv_psep_csi} one can readily write the PHSEP problem for imperfect CSI as  
\begin{align}
    &\min_{\boldsymbol{\Theta}\in \mathcal{M},P\in \mathbb{R}_+}  \ P \\
   &\displaystyle -\hspace{-0.2em}\sum_{\xi=1}^2 \hspace{-0.1em}\text{ln}\pc{\Phi\pc{\sqrt{\frac{P}{\sigma_w^2}}\pc{ \text{tr}\pc{\boldsymbol{\Theta}  \hat{\boldsymbol{H}}_{\xi,k} }\hspace{-0.2em}-\hspace{-0.2em}\sqrt{2N} \epsilon_k }} }\hspace{-0.2em} \leq \beta_k, \text{for} \ k\in \mathcal{K}, \notag
\end{align}
where $\hat{\boldsymbol{b}}_{\text{R},k}=r_k\hat{\boldsymbol{h}}_k^H,$ $\hat{\boldsymbol{b}}_{\text{I},k}=i_k\hat{\boldsymbol{h}}_k^H$,
\begin{align}
    \hat{\boldsymbol{H}}_{1,k}=\begin{bmatrix}
    \text{Re}\{\hat{\boldsymbol{b}}^T_{\text{R},k}\} \\
  - \text{Im}\{\hat{\boldsymbol{b}}^T_{\text{R},k}\} 
\end{bmatrix}^T, \quad \hat{\boldsymbol{H}}_{2,k}=\begin{bmatrix}
    \text{Im}\{\hat{\boldsymbol{b}}_{\text{I},k}^T\}\\
    \text{Re}\{\hat{\boldsymbol{b}}_{\text{I},k}^T\}
\end{bmatrix}^T.
\end{align}

\subsection{\textcolor{r1}{Derivation of the UBSEP-based Worst-Case Optimization}}
\begin{figure*}[t]
\begin{center}
\input{figures/Fig4}
\caption{Scenario: $K=2$, $N=15$, $\alpha_s=\alpha_\theta=4$, $\rho_k=10^{-\tau}$ for $k \in \mathcal{K}$, $P_\text{B}= 2$ dB and $\gamma=4$ dB. $P_n$ $\times$ $\tau$ (left),  $\overline{B}$ $\times$ $\tau$ (right).} 
\label{fig:power}       
\end{center}
\end{figure*}
\textcolor{r1}{The PUBSEP problem under imperfect CSI is written by applying the channel formulation with the UBSEP functions which yields
\begin{align}
    &\min_{\boldsymbol{\theta} \in \mathcal{T}^N,P\in \mathbb{R}_+}  \ P \\
    &\hspace{0.5em}\text{s.t.}\hspace{0.5em} \displaystyle \sum_{\xi=1}^2 \frac{1}{2}\text{erfc}\pc{\frac{\hat{d}_{\xi,k}\pc{\boldsymbol{\theta},P}+\tilde{d}_{\xi,k}\pc{\boldsymbol{\theta},P}}{\sigma_w}}\leq \rho_k, \ \text{for} \ k\in \mathcal{K}, \notag
\end{align}
where 
\begin{align}
    \hat{d}_{1,k}&=\sqrt{{P}}(\text{Re}\{s_k^*\hat{\boldsymbol{h}}_k^H \boldsymbol{\theta}\} \sin{\phi}-\text{Im}\{s_k^* \hat{\boldsymbol{h}}_k^H \boldsymbol{\theta}\} \cos{\phi}), \\ 
    \hat{d}_{2,k}&=\sqrt{P}(\text{Re}\{s_k^* \hat{\boldsymbol{h}}_k^H \boldsymbol{\theta}\} \sin{\phi}+\text{Im}\{s_k^*\hat{\boldsymbol{h}}_k^H \boldsymbol{\theta}\} \cos{\phi}),\\
    \tilde{d}_{1,k}&=\sqrt{{P}}(\text{Re}\{s_k^*\tilde{\boldsymbol{h}}_k^H \boldsymbol{\theta}\} \sin{\phi}-\text{Im}\{s_k^* \tilde{\boldsymbol{h}}_k^H \boldsymbol{\theta}\} \cos{\phi}),\\
    \tilde{d}_{2,k}&=\sqrt{P}(\text{Re}\{s_k^* \tilde{\boldsymbol{h}}_k^H \boldsymbol{\theta}\} \sin{\phi}+\text{Im}\{s_k^*\tilde{\boldsymbol{h}}_k^H \boldsymbol{\theta}\} \cos{\phi}).
\end{align}
For given $\boldsymbol{\theta}$ and $P$, the vector of $\tilde{\boldsymbol{h}}_{k}$ which yields the $k$-th user largest UBSEP (LUBSEP) is determined by solving the following optimization problem
\begin{align}
\label{opt:lubsep_ncov}
    &\text{LUBSEP}(\tilde{\boldsymbol{h}}_{k})=\max_{\tilde{\boldsymbol{h}}_{k}} \ \displaystyle \sum_{\xi=1}^2 \frac{1}{2}\text{erfc}\pc{\frac{\hat{d}_{\xi,k}+\tilde{d}_{\xi,k}\pc{\boldsymbol{\theta},P}}{\sigma_w}}\\
    &\hspace{0.5em}\text{s.t.}\hspace{0.5em} \|\tilde{\boldsymbol{h}}_{k}\|_2\leq {\epsilon_k}, \notag\\
    &\hspace{1.5em} \tilde{d}_{1,k}=\sqrt{{P}}\pc{\text{Re}\{s_k^*\tilde{\boldsymbol{h}}_k^H \boldsymbol{\theta}\} \sin{\phi}-\text{Im}\{s_k^* \tilde{\boldsymbol{h}}_k^H \boldsymbol{\theta}\} \cos{\phi}},\notag\\
    &\hspace{1.5em}\tilde{d}_{2,k}=\sqrt{P}\pc{\text{Re}\{s_k^* \tilde{\boldsymbol{h}}_k^H \boldsymbol{\theta}\} \sin{\phi}+\text{Im}\{s_k^*\tilde{\boldsymbol{h}}_k^H \boldsymbol{\theta}\} \cos{\phi}}\notag.
\end{align}
Problem \eqref{opt:lubsep_ncov} is only convex for $\hat{d}_{\xi,k}+\tilde{d}_{\xi,k}\pc{\boldsymbol{\theta},P}\leq0$ for $\xi\in\{1,2\}$, which is not a case of interest since it yields SEP greater than half. To avoid the computation of \eqref{opt:lubsep_ncov}, an upper bound on the $k$-th user LUBSEP is considered by independently maximizing $f_{\xi,k}(\tilde{\boldsymbol{h}}_{k})=\frac{1}{2}\text{erfc}\pc{\frac{\hat{d}_{\xi,k}+\tilde{d}_{\xi,k}(\tilde{\boldsymbol{h}}_{k})}{\sigma_w}}$, for $\xi\in\{1,2\}$. 
Since $\text{erfc}(\cdot)$ is a monotonically decreasing function $f_{\xi,k}(\cdot)$ is maximized when $\tilde{d}_{\xi,k}(\cdot)$ is minimized. Defining $\boldsymbol{\alpha}_k=s_k^*\boldsymbol{\theta}$, $\boldsymbol{\lambda}_{\text{R},k}=\text{Re}\{\boldsymbol{\alpha}_k\}$
, $\boldsymbol{\lambda}_{\text{I},k}=\text{Im}\{\boldsymbol{\alpha}_k\}$, 
\begin{align}
    \boldsymbol{\alpha}_{\text{R},k}&=[[\boldsymbol{\lambda}_{R,k}]_1, -[\boldsymbol{\lambda}_{I,k}]_1,\hdots,[\boldsymbol{\lambda}_{R,k}]_N, -[\boldsymbol{\lambda}_{I,k}]_N]^T\sin(\phi), \notag\\
    \boldsymbol{\alpha}_{\text{I},k}&=[\boldsymbol{\lambda}_{I,k}]_1\}, [\boldsymbol{\lambda}_{R,k}]_1,\hdots,\boldsymbol{\lambda}_{I,k}]_N\},[\boldsymbol{\lambda}_{R,k}]_N]^T\cos(\phi), \notag\\
    \tilde{\boldsymbol{h}}_{\text{r},k}&=[\text{Re}\{[\boldsymbol{h}_{k}^*]_1\}, \text{Im}\{[\boldsymbol{h}_{k}^*]_1\},\hdots,\text{Re}\{[\boldsymbol{h}_{k}^*]_N\},\text{Im}\{[\boldsymbol{h}_{k}^*]_N\}]^T,\notag
\end{align}
one can write $\tilde{d}_{1,k}$ and $\tilde{d}_{2,k}$ as
\begin{align}
\tilde{d}_{1,k}(\tilde{\boldsymbol{h}}_{\text{r},k})&=\sqrt{P}\tilde{\boldsymbol{h}}_{\text{r},k}^T(\boldsymbol{\alpha}_{\text{R},k}-\boldsymbol{\alpha}_{\text{I},k}),  \\
\tilde{d}_{2,k}(\tilde{\boldsymbol{h}}_{\text{r},k})&=\sqrt{P}\tilde{\boldsymbol{h}}_{\text{r},k}^T(\boldsymbol{\alpha}_{\text{R},k}+\boldsymbol{\alpha}_{\text{I},k}).
\end{align}
It becomes clear that $\tilde{d}_{1,k}(\tilde{\boldsymbol{h}}_{\text{r},k})$ and $\tilde{d}_{2,k}(\tilde{\boldsymbol{h}}_{\text{r},k})$ are minimized for $\tilde{\boldsymbol{h}}_{\text{r},k}= -\frac{\epsilon_k}{\sqrt{N}}(\boldsymbol{\alpha}_{\text{R},k}-\boldsymbol{\alpha}_{\text{I},k})$, and $\tilde{\boldsymbol{h}}_{\text{r},k}=-\frac{\epsilon_k}{\sqrt{N}}\pc{\boldsymbol{\alpha}_{\text{R},k}+\boldsymbol{\alpha}_{\text{I},k}}$, respectively.
Finally, the minimum values of $\tilde{d}_{\xi,k}$ read as $\tilde{d}_{\xi,k}=-\sqrt{P} \sqrt{N} \epsilon_k$, for $\xi\in\{1,2\}$. An upper bound on the UBSEP for imperfect CSI is then constructed as 
\begin{align}
    f_k(\boldsymbol{\theta})
    &=\displaystyle \sum_{\xi=1}^2 \frac{1}{2}\text{erfc}\pc{\frac{\hat{d}_{\xi,k}}{\sigma_w}-\sqrt{\frac{P}{\sigma_w^2}} \sqrt{N} \epsilon_k }.
\end{align}
Based on $f_k(\boldsymbol{\theta})$ the power minimization problem can be rewritten as 
\begin{align}
\label{opt:cv_pubsep_csi}
    &\min_{\boldsymbol{\theta} \in \mathcal{T}^N,P\in \mathbb{R}_+}  \ P \\
    &\text{s.t.}\hspace{0.5em} \sum_{\xi=1}^2 \frac{1}{2}\text{erfc}\pc{\frac{\hat{d}_{\xi,k}(\boldsymbol{\theta})}{\sigma_w}-\sqrt{\frac{P}{\sigma_w^2}} \sqrt{N} \epsilon_k } \leq \rho_k, \ \text{for} \ k\in \mathcal{K}.
    \notag
\end{align}
Note that \eqref{opt:cv_pubsep_csi} is convex for $\hat{d}_{\xi,k}(\boldsymbol{\theta})\geq \sqrt{P}\sqrt{N}\epsilon_k$, for $\xi\in\{1,2\}$ and $k\in\mathcal{K}$. With this, the proposed PUBSEP problem for imperfect CSI is assembled including the constraints $\hat{d}_{\xi,k}(\boldsymbol{\theta})\geq \sqrt{P}\sqrt{N}\epsilon_k$ and rewriting \eqref{opt:cv_pubsep_csi} with real-valued variables, which reads as 
\begin{align}
    &\min_{\boldsymbol{\theta}_\text{r},P\in \mathbb{R}_+}  \ P \\
    &\hspace{0.5em}\text{s.t.}\hspace{0.5em} \pr{\boldsymbol{\theta}_\text{r}}_{2n-1}+ j\pr{\boldsymbol{\theta}_\text{r}}_{2n} \in \mathcal{T}, \ \text{for} \ n\in \mathcal{N}, \ \ \hat{\boldsymbol{U}}\boldsymbol{\theta}_\text{r}\succeq \sqrt{N}\boldsymbol{\epsilon}, \notag \\
    &\hspace{0.5em} \displaystyle \sum_{\xi=1}^2 \frac{1}{2}\text{erfc}\pc{ \sqrt{\frac{P}{\sigma_w^2}} \pc{\hat{\boldsymbol{u}}_{\xi,k}^T \boldsymbol{\theta}_{\text{r}}-{\sqrt{N}\epsilon_k}} }\leq \rho_k, \ \text{for} \ k\in \mathcal{K}, \notag
\end{align}
where $\hat{\boldsymbol{u}}_{1,k}=\pc{\hat{\boldsymbol{\gamma}}_{\text{R},k}\sin\pc{\phi} -\hat{\boldsymbol{\gamma}}_{\text{I},k}\cos\pc{\phi}}^T$, $\hat{\boldsymbol{u}}_{2,k}=\pc{\hat{\boldsymbol{\gamma}}_{\text{R},k}\sin\pc{\phi} +\hat{\boldsymbol{\gamma}}_{\text{I},k}\cos\pc{\phi}}^T$, 
$\hat{\boldsymbol{\gamma}}_{\text{R},k}=
    [\textrm{Re}\{[\hat{\boldsymbol{\zeta}}_k]_{1}\}, -\textrm{Im}\{[\hat{\boldsymbol{\zeta}}_k]_{1}\},\hdots,\textrm{Re}\{[\hat{\boldsymbol{\zeta}}_k]_{N}\}, -\textrm{Im}\{[\hat{\boldsymbol{\zeta}}_k]_{N}\}]$, $\hat{\boldsymbol{\gamma}}_{\text{I},k}=
    [\textrm{Im}\{[\hat{\boldsymbol{\zeta}}_k]_{1}\}, \textrm{Re}\{[\hat{\boldsymbol{\zeta}}_k]_{1}\} ,\hdots,
\textrm{Im}\{[\hat{\boldsymbol{\zeta}}_k]_{N}\}, \textrm{Re}\{[\hat{\boldsymbol{\zeta}}_k]_{N}\}]$, $\tilde{\boldsymbol{\zeta}}_{k}=s_k^{*}\tilde{\boldsymbol{h}}_k^H$, $\boldsymbol{\epsilon}=[\epsilon_1,\hdots,\epsilon_K,\epsilon_1,\hdots,\epsilon_K]^T$.
Departing from \eqref{opt:cv_pubsep_csi} one can readily write the PHUBSEP problem for imperfect CSI as  
\begin{align}
    &\min_{\boldsymbol{\Theta}\in \mathcal{M},P\in \mathbb{R}_+}  \ P \\
   &\hspace{0.5em} \displaystyle \sum_{\xi=1}^2 \frac{1}{2}\text{erfc}\pc{ {\sqrt{\frac{P}{\sigma_w^2}}}\pc {\text{tr}\pc{\boldsymbol{\Theta}  \hat{\boldsymbol{U}}_{\xi,k} }- \sqrt{N} \epsilon_k } } \leq \rho_k, \ \text{for} \ k\in \mathcal{K}, \notag
\end{align}
where $\hat{\boldsymbol{a}}_k=s_k^*\hat{\boldsymbol{h}}_k^H$, and
\begin{align}
    \hat{\boldsymbol{U}}_{1,k}&=\begin{bmatrix}
        \text{Re}\{\hat{\boldsymbol{a}}^T_k\} \sin(\phi) - \text{Im}\{\hat{\boldsymbol{a}}^T_k\} \cos(\phi)\\
        -\text{Re}\{\hat{\boldsymbol{a}}^T_k\} \cos(\phi) - \text{Im}\{\hat{\boldsymbol{a}}^T_k\} \sin(\phi)
    \end{bmatrix}^T, \notag\\ 
    \hat{\boldsymbol{U}}_{2,k}&=\begin{bmatrix}
        \text{Re}\{\hat{\boldsymbol{a}}_k^T\} \sin(\phi) + \text{Im}\{\hat{\boldsymbol{a}}_k^T\} \cos(\phi)\\
        \text{Re}\{\hat{\boldsymbol{a}}_k^T\} \cos(\phi) - \text{Im}\{\hat{\boldsymbol{a}}_k^T\} \sin(\phi)
    \end{bmatrix}^T. \notag
\end{align}
}

\section{Numerical Results}
\label{sec:numerical_results_ris}

This section evaluates the proposed algorithms against the state-of-the-art approach from \cite{liu2021intelligent} in terms of UBCO and average normalized transmit power (ANTP) defined as $P_n=10 \log_{10}(\sfrac{P}{\sigma_w^2})$. For the simulations, the channel coefficients are modeled by independent Rayleigh fading, and the noise variance is considered to be $\sigma_w^2=1$. To simplify the analysis all users are considered to have the same SEP requirement such that ${\rho}_k=10^{-\tau}$, for $k\in\mathcal{K}$. A normalized target power budget $P_\text{B}=10 \log_{10}(\sfrac{P_\text{b}}{\sigma_w^2})$ [dB] is considered for the plots, with $P_\text{b}$ being the target power budget in linear scale utilized in the proposed PBB algorithms. \textcolor{r1}{The evaluated approaches considered are summarized in table \ref{tab:table1}.}
\begin{table*}
\centering
{\scriptsize{
  \caption{{\textcolor{r1}{UBCO of the Algorithms}} }
  \label{tab:table1}
  \begin{tabular}{ | p {10em}  | p {27em} | p{15em}| p{10em}|}
  \hline
    {\textcolor{r1}{Abbreviation}} & \textcolor{r1}{Meaning} & \textcolor{r1}{{Reference}} &\textcolor{r1}{UBCO} \\ \hline \hline
    \textcolor{r1}{Proposed PSEP PBB}&  \textcolor{r1}{Power Min. under SEP const. via Partial Branch-and-Bound}& \textcolor{r1}{Apply Algorithm \ref{alg:bnb} to \eqref{opt:pm_sep}}  &\textcolor{r1}{{$\mathcal{O}({B (N^{3.5}+N^3\sqrt{K})})$}}\\ \hline
    \textcolor{r1}{Proposed PUBSEP PBB}&  \textcolor{r1}{Power Min. under UBSEP const. via Partial Branch-and-Bound}& \textcolor{r1}{Apply Algorithm \ref{alg:bnb} to \eqref{opt:rv_slp}} &\textcolor{r1}{{$\mathcal{O}({B (N^{3.5}+N^3\sqrt{K})})$}}\\ \hline
    \textcolor{r1}{PSEP FBB}&  \textcolor{r1}{Power Min. under UBSEP const. via Full Branch-and-Bound} & \textcolor{r1}{Apply FBB to \eqref{opt:pm_sep}}&\textcolor{r1}{{$\mathcal{O}({B (N^{3.5}+N^3\sqrt{K})})$}}\\ \hline
    \textcolor{r1}{PUBSEP FBB}&  \textcolor{r1}{Power Min. under UBSEP const. via Full Branch-and-Bound} & \textcolor{r1}{Apply FBB to \eqref{opt:rv_slp}}&\textcolor{r1}{{$\mathcal{O}({B (N^{3.5}+N^3\sqrt{K})})$}}\\ \hline
    \textcolor{r1}{Proposed PHSEP}&  \textcolor{r1}{Power Min. under SEP const. for High-Resolution RIS } & \textcolor{r1}{Apply Algorithm \ref{alg:bisection}  to \eqref{opt:pmhr_sep}}&\textcolor{r1}{{$\mathcal{O}(N^{2.5})$}}\\ \hline
    \textcolor{r1}{Proposed PHUBSEP}&  \textcolor{r1}{Power Min. under UBSEP const. for High-Resolution RIS } & \textcolor{r1}{Apply Algorithm \ref{alg:bisection}  to \eqref{opt:rv_slp_unconstrained}}&\textcolor{r1}{{$\mathcal{O}(N^{2.5})$}}\\ \hline
    \textcolor{r1}{PHMMDDT \cite{liu2021intelligent}}&  \textcolor{r1}{Power Min. under MMDDT const. for High-Resolution RIS} & \textcolor{r1}{Apply RCG to \eqref{opt:mmddt_ris} cf. \cite{liu2021intelligent}}&\textcolor{r1}{{$\mathcal{O}(N^{2.5})$}}\\ \hline
  \end{tabular}
}}\end{table*}


\subsection{Performance Analysis versus SEP requirement}
\begin{figure*}[ht]
\begin{center}
\input{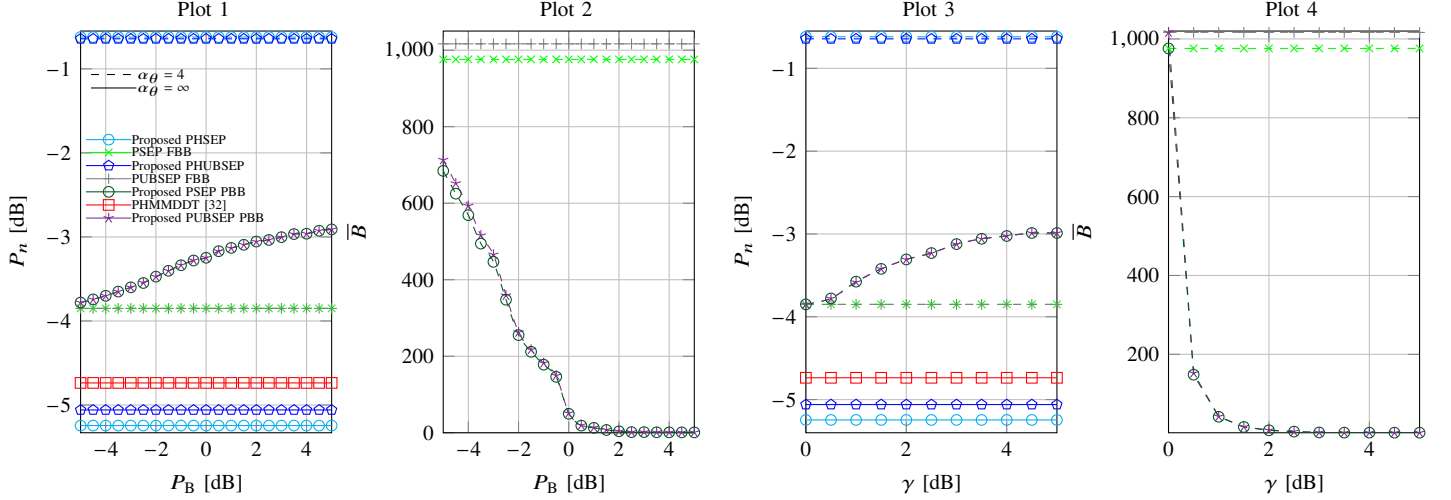}
\caption{\textcolor{r1}{Scenario: $K=2$, $N=15$, $\alpha_s=\alpha_\theta=4$, $\rho_k=10^{-4}$ for $k \in \mathcal{K}$. $P_n$ [dB] $\times$ $P_\text{B}$ (Plot 1), $\overline{B}$ $\times$ $P_\text{B}$ (Plot 2), for $\gamma=0$ dB. $P_n$ [dB] $\times$ $\gamma$ (Plot 3), $\overline{B}$ $\times$ $\gamma$ (Plot 4), for $P_\text{B}=-\infty$ dB.}}
\label{fig:trade_offs}       
\end{center}
\end{figure*}
This section considers a scenario with $K=2$ users, $N=15$ reflecting elements, and QPSK data and transmit symbols, i.e., $\alpha_s=\alpha_\theta=4$. For the experiments of Fig.~\ref{fig:power} the proposed PBB approaches utilize $P_\text{B}= 2$ dB and $\gamma=4$ dB.
The first experiment consists of the evaluation of the ANTP required for attaining the SEP requisites versus the SEP requirement parameter $\tau$. The left-hand side (LHS) of Fig.~\ref{fig:power} shows that the unquantized PHSEP approach outperforms the unquantized PHUBSEP method in terms of ANTP for all values of $\tau$. It also shows that the proposed unquantized PHSEP and PHUBSEP techniques require smaller ANTP for attaining the SEP requisites than the PHMMDDT formulation from \cite{liu2021intelligent} for all values of $\tau$. Note that, although not guaranteed due to the suboptimality of the BM utilized for solving the problems, this is expected since, as stated in section \ref{sec:formulation}, the PUBSEP formulation is a restriction of the PSEP optimization problem, and, as shown in appendix \ref{app:mmddt}, the PHMMDDT formulation is a restriction of the PHUBSEP problem. Regarding the finite resolution methods, the LHS of Fig.~\ref{fig:power} shows that the proposed PSEP and PUBSEP PBB methods and the PSEP and PUBSEP FBB approaches yield a significant decrease in required ANTP compared to the high-resolution approaches after quantization. Moreover, it \textcolor{r3}{is} seen that the proposed PSEP and PUBSEP PBB and FBB methods yield approximately $2$ dB and $1.2$ dB loss in relation to the infinite resolution approaches, respectively.

The UBCO analysis is done considering that the convex optimization-based approaches are solved with the barrier method. The UBCO of considered approaches is shown in Table~\ref{tab:table1}, where $B$ denotes the given number of subproblems solved in the corresponding branch-and-bound algorithm. As shown in Table~\ref{tab:table1} the high-resolution approaches yield significantly smaller UBCO than the branch-and-bound approaches which justifies its utilization for scenarios where the resolution of the RIS elements is sufficiently high such that the decrease in ANTP performance is relatively small. These scenarios are explored in section \ref{sec:resolution}. Since the UBCO of the branch-and-bound approaches depends on the value of $B$, for comparing their complexity an evaluation of $B$ is necessary. This evaluation is done in the second experiment in terms of the average value of $B$, termed $\overline{B}$, and is shown in the right-hand side (RHS) of Fig.~\ref{fig:power}. 

The RHS of Fig.~\ref{fig:power} shows that the PSEP designs yield reduced $\overline{B}$ when compared with PUBSEP, with this it is concluded that PSEP is favorable in terms of UBCO when compared with PUBSEP. Moreover, the RHS of Fig.~\ref{fig:power} shows a complexity reduction of at least factor $540$ when utilizing the proposed PBB approaches compared with its FBB counterparts. Finally, a joint analysis of the plots in Fig.~\ref{fig:power}  summarizes the complexity-performance trade-off achieved when utilizing the $P_\text{B} = 2$ dB and $\gamma$ = 4 dB. Fig.~\ref{fig:power} shows that using the proposed PBB yields a power increase smaller than 1 dB and a UBCO decrease of at least factor 540 compared to FBB counterparts.

\subsection{Performance-Complexity Trade-off Evaluation of the Proposed Branch-and-Bound Methods}
\label{sec:trade_off}
\begin{figure*}[t]
\begin{center}
\input{figures/Fig6}
\caption{\textcolor{r1}{Scenario: $N=30$, $\alpha_s=4$, $\rho_k=10^{-3}$ for $k \in \mathcal{K}$, $P_\text{B}=3$ dB, $\gamma=2$ dB. $P_n$ [dB] $\times$ $K$ (left), $\overline{B}$ $\times$ $K$ (right)}} 
\label{fig:users}       
\end{center}
\end{figure*}
The performance analysis of the proposed PBB approaches presented in Fig.~\ref{fig:power} is computed considering the $P_\text{B}= 2$ dB and $\gamma=4$ dB, which corresponds to the specific complexity performance trade-off shown. Yet, by varying the values of $P_\text{B}$ and $\gamma$ different trade-offs are achievable. This section evaluates the performance of the proposed PBB approaches for the different values of the acceptable power increase $\gamma$ and normalized target power budget $P_\text{B}$. For this section, the SEP requirements are set to $\rho_k=10^{-4}$ for $k \in \mathcal{K}$.

\textcolor{r1}{The first experiment, shown in the first two plots of Fig.~\ref{fig:trade_offs}, evaluates the impact of the target power budget $P_\text{B}$ in the performance of the different approaches, for no acceptable power increase, meaning $\gamma=0$ dB. As shown in Plot 1 of Fig.~\ref{fig:trade_offs}, for the low-end values of $P_\text{B}$ the ANTP performance of the proposed PSEP and PUBSEP PBB methods is similar to its FBB counterparts. This is the case since for extremely restrictive power budget scenarios where attaining $P_\text{out}\leq P_\text{B}$ is not possible (or is possible only with the optimal solution) the proposed PBB approach when operating with $\gamma=0$ dB, yields the optimal solution with FBB complexity. Yet, as $P_\text{B}$ increases, the target power budget of the system starts to be achieved with suboptimal reduced complexity reflection coefficients, and thus, the ANTP performance of the proposed PBB approaches starts to deviate from the performance of the FBB methods. Plot 2 of Fig.~\ref{fig:trade_offs} further highlights this behavior where it can be seen that, as the target power budget of the system increases, the average number of evaluated bounds explored by the proposed PSEP and PUBSEP PBB approaches decrease until, for $P_\text{B}=3$ dB, $\overline{B}\approx 1$ which implies a UBCO of $\mathcal{O}(N^{3.5}+N^3 \sqrt{K})$.
In the second experiment, shown in plots 3 and 4 of Fig.~\ref{fig:trade_offs}, the performance of the proposed PBB is evaluated for different values of $\gamma$ considering $P_\text{B}=-\infty$ dB. Plots 3 and 4 of Fig.~\ref{fig:trade_offs} show, as expected, that for $\gamma=0$ dB the proposed PSEP and PUBSEP PBB approaches yield the same ANTP and average number of subproblems solved as its FBB counterparts. Plot 3 of Fig.~\ref{fig:trade_offs} shows that as $\gamma$ increases, i.e., as the system accepts a larger power increase, the ANTP of the proposed PBB approaches grows. Note, however, that the real increase in ANTP, shown in Plot 3 of Fig.~\ref{fig:trade_offs}, is always significantly smaller than the acceptable power increase being on average approximately $17\%$ of the value of $\gamma$. Plot 4 of Fig.~\ref{fig:trade_offs} on the other hand shows a rapid decrease in $\overline{B}$ with an increase in $\gamma$. This underlines the idea that, although computing the optimal solution requires exploration of a large number of branches when allowing for small ANTP compromises one can achieve significant reduction such that the resulting UBCO is $\mathcal{O}(N^{3.5}+N^3 \sqrt{K})$. 
Finally, a joint analysis of the plots of Fig.~\ref{fig:trade_offs} illustrates that a decrease in UBCO by a factor greater than $100$ can be achieved with an increase of ANTP of less than $0.45$ dB. With this, it can be stated that significant complexity reduction can be achieved with minor ANTP compromise.}

\subsection{\textcolor{r1}{Performance Analysis versus Number of Users}}
\label{sec:users}

\textcolor{r1}{The proposed formulations implicitly consider settings in which the degrees of freedom of the system are sufficient to allow for scenarios in which problem \eqref{opt:slp_original_0} has a feasible solution. In practice, due to the RIS' number of elements and available phase shifts being finite, only a finite number of users can be served simultaneously. This section evaluates how the performance of the considered approaches is affected by the growth in the number of users and how many users the proposed techniques can serve before \eqref{opt:slp_original_0} becomes infeasible with the given parameters. The values of $P_\text{B}=0$ dB and $\gamma=3$ are considered for the experiments in this section. Moreover, the SEP requisites are set to $\rho_k=10^{-3}$ for $k \in \mathcal{K}$, and $N=
30$ reflecting elements are considered.}

\textcolor{r1}{The first experiment, present in the LHS of Fig.~\ref{fig:users}, evaluates the ANTP required to attain the SEP requisites for a given number of users. As expected the LHS of Fig.~\ref{fig:users} shows that the infinite resolution approaches support serving all evaluated numbers of users with the smallest ANTP. Moreover, it is seen that the proposed high-resolution approaches outperform the PHMMDDT baseline from \cite{liu2021intelligent}. Fig.~\ref{fig:users} shows that all proposed approaches can support serving $K=8$ users with $\alpha_\theta=8$ available phase-shifts and that the proposed PBB approaches require reduced transmit power when compared with the proposed PHSEP and PHUBSEP  techniques. Regarding the case of $\alpha_\theta=4$, while the proposed PHSEP and PHUBSEP techniques support serving only $K=4$ users, the proposed PSEP and PUBSEP PBB methods can simultaneously serve $K=6$ users. Finally, the second experiment, shown in the RHS of Fig.~\ref{fig:users}, evaluates $\overline{B}$ for the different values of $K$. As expected the complexity of the proposed PBB techniques grows with the number of users served. This is the case since, as $K$ increases, it becomes more difficult to attend to the target power budget of the system and the PBB approach starts requiring to evaluate more branches of the tree to compute a solution.}

\subsection{Performance Analysis versus Resolution}
\label{sec:resolution}
\begin{figure}[t]
\begin{center}
%
%
%
\usetikzlibrary{positioning,calc}

\definecolor{mycolor1}{rgb}{0.00000,1.00000,1.00000}%
\definecolor{mycolor2}{rgb}{1.00000,0.00000,1.00000}%

\pgfplotsset{every axis label/.append style={font=\footnotesize},
every tick label/.append style={font=\footnotesize}
}

\hspace{-6em}

\begin{tikzpicture}[spy using outlines={rectangle,magnification=3,connect spies}] 
\begin{axis}[%
name=A,
ymode=linear,
width  = 0.85\columnwidth,
height = 0.60\columnwidth,
scale only axis,
xmin  = 2,
xmax  = 7,
xlabel= {$b$},
xmajorgrids,
ymin=-5.5,
ymax=-0.5,
ylabel={$P_n$ [dB] },
ymajorgrids,
legend entries={Proposed PHSEP,
Proposed PHUBSEP,
PHMMDDT \cite{liu2021intelligent},
Proposed PSEP PBB,
Proposed PUBSEP PBB
},
                legend columns=1,
legend style={at={(1,1)},anchor=north east ,draw=none,fill=none,legend cell align=left,font=\tiny}
]






\addplot+[smooth,color=cyan,solid, every mark/.append style={solid, fill=cyan!50},mark=o,
y filter/.code={\pgfmathparse{\pgfmathresult-0}\pgfmathresult}]
  table[row sep=crcr]{%
     2      -5.24503823635966    \\
     3      -5.24503823635966    \\
     4      -5.24503823635966     \\
     5      -5.24503823635966    \\
     6      -5.24503823635966   \\
     7      -5.24503823635966   \\
};

\addplot+[smooth,color=blue,solid, every mark/.append style={solid, fill=cyan!50},mark=pentagon,
y filter/.code={\pgfmathparse{\pgfmathresult-0}\pgfmathresult}]
  table[row sep=crcr]{%
     2    -5.02656774860919    \\
     3    -5.02656774860919    \\
     4    -5.02656774860919    \\
     5    -5.02656774860919    \\
     6    -5.02656774860919      \\
     7    -5.02656774860919    \\
};

\addplot+[smooth,color=red,solid, every mark/.append style={solid, fill=cyan!50},mark=square,
y filter/.code={\pgfmathparse{\pgfmathresult-0}\pgfmathresult}]
  table[row sep=crcr]{%
     2    -3.60765844920452     \\
     3    -3.60765844920452     \\
     4    -3.60765844920452     \\
     5    -3.60765844920452     \\
     6    -3.60765844920452      \\
     7    -3.60765844920452     \\
};

\addplot+[smooth,color=dark_green,solid, every mark/.append style={solid, fill=cyan!50},mark=o,
y filter/.code={\pgfmathparse{\pgfmathresult-0}\pgfmathresult}]
  table[row sep=crcr]{%
0 0\\
};

\addplot+[smooth,color=purple,solid, every mark/.append style={solid, fill=cyan!50},mark=star,
y filter/.code={\pgfmathparse{\pgfmathresult-0}\pgfmathresult}]
  table[row sep=crcr]{%
0 0\\
};

\addplot+[smooth,color=dark_green,dashed, every mark/.append style={solid, fill=cyan!50},mark=o,
y filter/.code={\pgfmathparse{\pgfmathresult-0}\pgfmathresult}]
  table[row sep=crcr]{%
     2    -3.15054945713062     \\
     3    -4.52941503626069     \\
     4    -4.92726191680335     \\
     5    -5.08295935399363     \\
     6    -5.12527444263284      \\
     7    -5.13902185503343     \\
};

\addplot+[smooth,color=purple,dashed, every mark/.append style={solid, fill=cyan!50},mark=star,
y filter/.code={\pgfmathparse{\pgfmathresult-0}\pgfmathresult}]
  table[row sep=crcr]{%
     2    -3.15054508753892   \\
     3    -4.52940775415115   \\
     4    -4.92725254622411   \\
     5    -5.08250097484720   \\
     6    -5.12732347606802     \\
     7    -5.12506895267494   \\
};

\addplot+[smooth,color=cyan,dashed, every mark/.append style={solid, fill=cyan!50},mark=o,
y filter/.code={\pgfmathparse{\pgfmathresult-0}\pgfmathresult}]
  table[row sep=crcr]{%
     2    -0.644399375205388        \\
     3    -4.13998422826567     \\
     4    -4.85746694202900     \\
     5    -5.14011812348281     \\
     6    -5.21189315259226       \\
     7    -5.23769547317812     \\
};

\addplot+[smooth,color=blue,dashed, every mark/.append style={solid, fill=cyan!50},mark=pentagon,
y filter/.code={\pgfmathparse{\pgfmathresult-0}\pgfmathresult}]
  table[row sep=crcr]{%
     2    -0.666521318208423    \\
     3    -4.04347987964919    \\
     4    -4.76134358820731    \\
     5    -4.95343789155766    \\
     6    -5.00180712717085      \\
     7    -5.01977810378150    \\
};

\addplot[smooth,color=black,dashed,mark=no_mark
y filter/.code={\pgfmathparse{\pgfmathresult-0}\pgfmathresult}]
  table[row sep=crcr]{%
	1 2\\
};\label{plot:alpha4}

\addplot[smooth,color=black,solid,
y filter/.code={\pgfmathparse{\pgfmathresult-0}\pgfmathresult}]
  table[row sep=crcr]{%
	1 2\\
};\label{plot:alpha_infty}

\node [draw=none,fill=none,font=\tiny,anchor= north  east] at (axis cs: 5.045,-0.5) {
\setlength{\tabcolsep}{0.5mm}
\renewcommand{\arraystretch}{.8}
\begin{tabular}{l}

\ref{plot:alpha4}{\hspace{0.5em}$\alpha_\theta=2^b$} \\
\ref{plot:alpha_infty}{\hspace{0.5em}$\alpha_\theta=\infty$}\\

\end{tabular}
};

\end{axis}

\end{tikzpicture}%
\caption{{Scenario: $K=2$, $N=15$, $\alpha_s=4$, $\rho_k=10^{-4}$ for $k \in \mathcal{K}$, $P_\text{B}=0$ dB and $\gamma=1$ dB. $P_n$ $\times$ $b$.}} 
\label{fig:resolution}       
\end{center}
\end{figure}
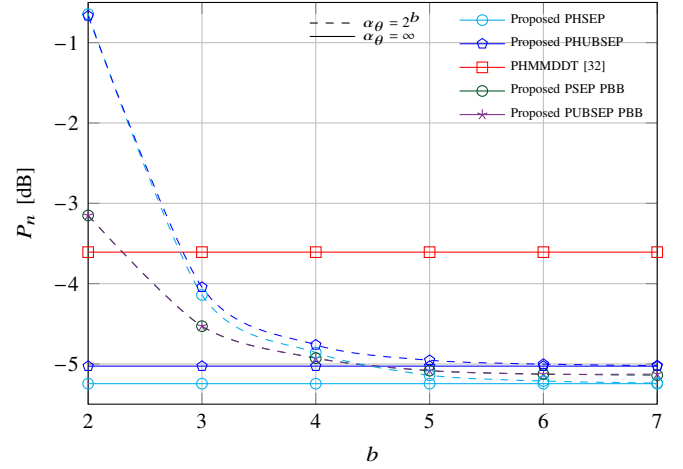
\begin{figure*}[t]
\begin{center}
\input{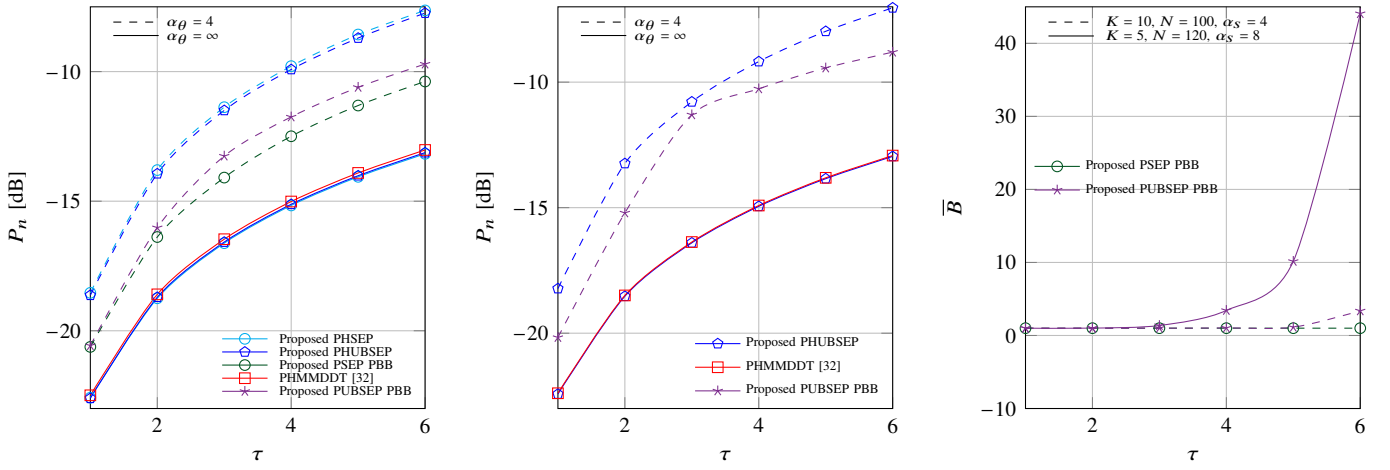}
\caption{\textcolor{r1}{Scenario: $\rho_k=10^{-\tau}$ for $k \in \mathcal{K}$, $\alpha_\theta=4$, $P_\text{B}= -8$ dB and $\gamma=1$ dB. $P_n\times\tau$, for $K=10$, $N=100$, $\alpha_s$ (left). $P_n\times\tau$, for $K=5$, $N=120$, $\alpha_s=8$ (center). $\overline{B}\times\tau$, both previous scenarios (right).} }
\label{fig:large_scale}       
\end{center}
\end{figure*}
\begin{figure*}[t]
\begin{center}
\input{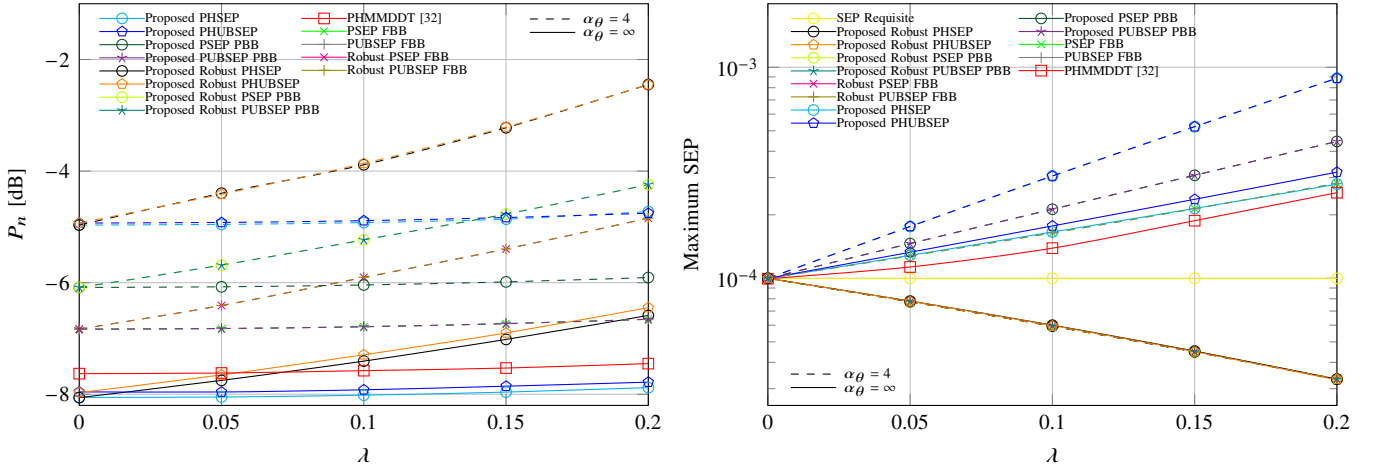}
\caption{\textcolor{r1}{Scenario: $K=2$, $N=20$, $\alpha_s=\alpha_\theta=4$, $\epsilon_k=\lambda$, $\rho_k=10^{-4}$ for $k\in\mathcal{K}$, $P_\text{B}= 0$ dB and $\gamma=1$ dB. $P_n$ $\times$ $\lambda$ (left). Maximum SEP $\times$ $\lambda$ (right). } }
\label{fig:CSI}       
\end{center}
\end{figure*}

This section evaluates the effects of the resolution of the RIS elements, measured in bits as $b=\log_2(\alpha_\theta)$, on the performance of the proposed methods. The scenario consists of a system with $K=2$ users, $N=15$ reflecting elements, QPSK data symbols, $\rho_k=10^{-4}$ for $k \in \mathcal{K}$, target power budget $P_\text{B}=0$ dB and $\gamma=1$ dB. The experiment consists of an evaluation of $P_n$ required for attaining the SEP constraints for different values of $b$. Fig.~\ref{fig:resolution} shows that, for $b=3$, all proposed methods outperform the infinite resolution PHMMDDT baseline from \cite{liu2021intelligent}. 
Moreover, as expected, Fig.~\ref{fig:resolution} shows that as the number of resolution bits increases the proposed PHSEP and PHUBSEP quantized methods approach the ANTP performance of their infinite resolution counterparts. Considering that the proposed PHSEP and PHUBSEP quantized techniques yield UBCO of $\mathcal{O}(N^{2.5})$ one can understand that the proposed PHSEP and PHUBSEP methods yield reduced ANTP with low complexity which highlights the efficiency of the proposed methods. Regarding the proposed PSEP and PUBSEP PBB methods Fig.~\ref{fig:resolution} shows that for $b\leq4$ the proposed PBB techniques yield the smallest ANTP of the quantized approaches. Yet, for $b>4$, they are outperformed by the PHSEP approach. Although the PHSEP approach yields a suboptimal solution even for $\alpha_\theta=\infty$, this is expected since $P_\text{B}=0$ dB and $\gamma=1$ dB are considered, and, with this, complexity reduction is achieved at the expense of ANTP performance. Regarding the UBCO of the proposed PBB approaches, the performance shown in Fig.~\ref{fig:resolution} is achieved with $1<\overline{B}<2.5$ for all $b$.

\subsection{Transmit Power Analysis for Large-Scale Systems}
\label{sec:large_scale}

This section evaluates the ANTP performance of the considered approaches for different values of the SEP requirement parameter $\tau$, in a massive MIMO system. Unlike FBB approaches, which can yield prohibitive complexity, the proposed PBB method is suitable for large-scale MIMO. For this section, the RIS elements are considered to have 2-bit resolution, and the PBB parameters are set to $P_\text{B}= -8$ dB and $\gamma=1$ dB.

\textcolor{r1}{The first experiment, shown in the LHS of Fig.~\ref{fig:large_scale}, considers a MIMO scenario of $K=10$ users, $N=100$ reflecting elements, and QPSK users' data. As shown the proposed PSEP and PUBSEP PBB techniques outperform the methods PHSEP and PHUBSEP methods after quantization, which highlights the suitability of the proposed PBB technique for large-scale MIMO. Regarding the infinite resolution approaches, the LHS of Fig.~\ref{fig:large_scale} shows that the proposed PHSEP and PHUBSEP techniques yield smaller ANTP when compared with the PHMMDDT technique. The second experiment, present in the center plot of Fig.~\ref{fig:large_scale}, considers a MIMO scenario of $K=5$ users, $N=120$ reflecting elements, and 8-PSK users' data. As shown the proposed PUBSEP PBB approach outperforms the quantized PHUBSEP method for all values of $\tau$. Yet, due to the small number of resolution bits of the RIS elements, the PUBSEP PBB approach yields approximately a $4$ dB increase in ANTP compared to the unquantized PHUBSEP method.
Finally, regarding the complexity of the proposed PBB approaches, the RHS of Fig.~\ref{fig:large_scale} shows the value of $\overline{B}$ required for achieving the results of the other plots. As seen in the RHS of Fig.~\ref{fig:large_scale} the results were achieved with the maximum values of $\overline{B}=3.37$ and $\overline{B}=44$ for $\tau=6$, respectively.
}


\subsection{\textcolor{r1}{Performance Evaluation for Imperfect CSI}}

\textcolor{r1}{This section evaluates the considered techniques for the case of imperfect CSI. The ellipsoid channel mismatch model from section \ref{sec:csi} is considered, and, to facilitate the analysis, all users are considered to have the same CSI quality, meaning $\sigma_{h,k}=\lambda$, for $k\in\mathcal{K}$. It is also considered that $\epsilon_k=\sigma_{h,k}$, and $\text{E}\{\|{\boldsymbol{h}}_k\|_2\}=1$, for $k\in\mathcal{K}$. For the experiments, $3\cdot10^5$ different realizations of $\boldsymbol{h}_k$ were considered. The LHS of Fig.~\ref{fig:CSI} shows an evaluation of the average value of $P_n$ versus $\lambda$. As expected, the robust techniques yield the same required $P_n$ as their non-robust counterparts for $\lambda=0$. Yet, as $\lambda$ increases the robust technique's value of $P_n$ increases, while its non-robust counterparts do not significantly vary. This is the case since, once the non-robust approaches do not consider CSI imperfection in their formulation, they do not adapt to the CSI mismatch level. The consequence is presented in the RHS of Fig.~\ref{fig:CSI} which shows the maximum SEP value attained with the different realizations versus $\lambda$. It is seen that, while the robust methods guarantee attainment of the SEP requisites for all cases, the SEP of the non-robust techniques can be higher than the requisite and thus does not necessarily meet the requisites of the system. 
}



\section{Conclusions}
\label{sec:conclusion}
This study considers a virtual multiuser MIMO system with PSK modulation realized via the RIS passive transmitter setup. With a discrete phase-shift RIS model the study proposes a power minimization problem under QoS constraints. The problem formulation is divided into two scenarios. First, for the case of QPSK user's data, the SEP is utilized as the QoS criterion. Then, for the general case of $M$-PSK, the UBSEP is used for designing the QoS constraints. Based on the considered formulations a PBB approach is proposed which allows solutions that either attain the target power budget of the system or are sufficiently close to the optimal such that further computation is considered unnecessary.  
For the special case of high-resolution RIS, the study adapts the PSEP and PUBSEP formulations based on the approximation of the discrete phase shift set by its continuous counterpart. The previously proposed problems are reformulated as constrained optimizations on an oblique manifold, and solved via the proposed BM. 
Numerical results underline that by utilizing the proposed PBB method significant complexity reduction can be achieved with negligible increase in transmit power. Moreover, numerical results show that for the case of high-resolution RIS the proposed techniques yield low complexity with high ANTP performance.

\appendix
\subsection{\textcolor{r3}{Convexity of the SEP Constraint Functions}}
\label{app:sep}

For proving convexity we depart from a function with known properties and apply a series of operations to arrive at $f_k(\boldsymbol{\theta}_\text{r})=-\sum_{\xi=1}^2\ln\pc{\Phi\pc{\sqrt{\frac{P}{\sigma_w^2}}\boldsymbol{h}_{\xi,k}^T\boldsymbol{\theta}_\text{r}}} - \beta_k$. To this end, consider the log-concave function \cite[Example 3.39]{Boyd_2004}, $\Phi(\theta)=\int_{-\infty}^\theta e^{-\frac{u^2}{2} }du$. A function $f$ is log-concave if $\log f$ is concave \cite[Definition 3.5.1]{Boyd_2004}. With this, $g(\theta)=\ln(\Phi(\theta))$ is a concave function. Note that, $h(\boldsymbol{\theta}_\text{r}) =g(\boldsymbol{A}\boldsymbol{\theta}_\text{r}+\boldsymbol{b})$ is concave if $g({\theta})$ is concave \cite[Section 3.2.2]{Boyd_2004}. With this, it follows that $h_1(\boldsymbol{\theta}_\text{r})=\ln(\Phi(\boldsymbol{h}_{1,k}\boldsymbol{\theta}_\text{r}))$ and $h_2(\boldsymbol{\theta}_\text{r})=\ln(\Phi(\boldsymbol{h}_{2,k}\boldsymbol{\theta}_\text{r}))$ are concave functions. As stated in \cite[Section 3.2.1]{Boyd_2004}, if $w_i\geq0$ and $f_i(\boldsymbol{\theta}_\text{r})$ is concave for all $i$, then $u(\boldsymbol{\theta}_\text{r})=\sum_i w_i f_i(\boldsymbol{\theta}_\text{r})$ is concave. By setting $w_i=1$ for $i\in\{1,2,3\}$ and $f_1(\boldsymbol{\theta}_\text{r})=\ln(\Phi(\boldsymbol{h}_{1,k}\boldsymbol{\theta}_\text{r}))$, $f_2(\boldsymbol{\theta}_\text{r})=\ln(\Phi(\boldsymbol{h}_{2,k}\boldsymbol{\theta}_\text{r}))$ and $f_3(\boldsymbol{\theta}_\text{r})=\beta_k$ one can assemble the concave function $u(\boldsymbol{\theta}_\text{r})=\sum_{\xi=1}^{2}\ln(\Phi(\boldsymbol{h}_{\xi,k}\boldsymbol{\theta}_\text{r}))+\beta_k$.
Note that $f_k(\boldsymbol{\theta}_\text{r})$ is convex since $f_k(\boldsymbol{\theta}_\text{r})=-u(\boldsymbol{\theta}_\text{r})$ and $u(\boldsymbol{\theta}_\text{r})$ is concave.
\subsection{Condition for convexity of the Union-Bound SEP functions}
\label{app:ubsep}
This section derives the conditions in which the UBSEP functions $f_k(\boldsymbol{x})=\frac{1}{2}\sum_{\xi=1}^2 \text{erfc}(\boldsymbol{\nu}_{\xi,k}^T \boldsymbol{x} )$ are convex. As stated in \cite[Section 3.1.4]{Boyd_2004} convexity can be proven by evaluating the conditions under which the Hessian is positive semi-definite (PSD). Taking the derivative of $f_k(\boldsymbol{x})$ with respect to $\boldsymbol{x}$ yields
\begin{align}
    \frac{\partial f_k(\boldsymbol{x})}{\partial \boldsymbol{x}}=-&\frac{1}{\sqrt{\pi}} \sum_{\xi=1}^2 \ e^{{-\pc{ \boldsymbol{v}_{\xi,k}^T  \boldsymbol{x}}^2} }\boldsymbol{v}_{\xi,k}. 
\end{align}
The Hessian is then computed by taking the derivative with respect to $\boldsymbol{x}^T$, which yields 
\begin{align}
\label{eq:Hess_theta}
   \frac{\partial f_k^2(\boldsymbol{x})}{\partial \boldsymbol{x}\partial \boldsymbol{x}^T}=&\frac{2}{\sqrt{\pi}}\sum_{\xi=1}^2 e^{{-\pc{\boldsymbol{v}_{\xi,k}^T  \boldsymbol{x}}^2} }\boldsymbol{v}_{\xi,k}\pc{\boldsymbol{v}_{\xi,k}^T  \boldsymbol{x}}\boldsymbol{v}_{\xi,k}^T.
\end{align}
Note that, a sufficient condition for $\nabla^2 f_k(\boldsymbol{x})$ to be PSD is $\boldsymbol{v}_{1,k}^T  \boldsymbol{x}\geq0$ and $\boldsymbol{v}_{2,k}^T  \boldsymbol{x}\geq0$. 

\subsection{Relation Between the MMDDT and UBSEP problems}
\label{app:mmddt}
When considering equation \eqref{eq:ubsep} one can construct an upper bound on $\text{P}_\text{ub}(\hat{s}_k|\boldsymbol{\theta},P)$ as
\begin{align}
\label{eq:mddt_bound}
    \text{P}_\text{ub}(\hat{s}_k|\boldsymbol{\theta},P)&=\frac{1}{2}\text{erfc}\pc{\frac{d_{1,k}\pc{\boldsymbol{\theta},P}}{\sigma_w}} +\frac{1}{2}\text{erfc}\pc{\frac{d_{2,k}\pc{\boldsymbol{\theta},P}}{\sigma_w}}\notag\\
    &\leq \text{erfc}\pc{ \frac{d_{k}\pc{\boldsymbol{\theta},P}}{\sigma_w}}, 
\end{align}
where $d_k\pc{\boldsymbol{\theta},P}$ is the MDDT utilized in \cite{liu2021intelligent} given by
    $d_k\pc{\boldsymbol{\theta},P}=\displaystyle \min_{\xi\in{1,2}} d_{\xi,k}\pc{\boldsymbol{\theta},P}=\sqrt{{P}}(\text{Re}\{s_k^*\boldsymbol{h}_k^H \boldsymbol{\theta}\} \sin{\phi}-\PM{\text{Im}\{s_k^* \boldsymbol{h}_k^H \boldsymbol{\theta}\}} \cos{\phi})$. 
One can write the high-resolution RIS power minimization problem with bound from \eqref{eq:mddt_bound} as 
\begin{align}
\label{opt:mmddt_problem1}
    &\min_{\boldsymbol{\theta}, P}  \ P \\
    &\hspace{0.7em}\text{s.t.}\hspace{0.5em} \PM{\pr{\boldsymbol{\theta}}_n}^2=1,\ \text{for} \ n \in \mathcal{N}, \hspace{1em} P\geq 0, \notag \\
    &\hspace{0.7em} \text{erfc}\pc{\frac{d_{k}\pc{\boldsymbol{\theta},P}}{\sigma_w}} \leq \rho_k, \ \text{for} \ k\in \mathcal{K}.\notag 
\end{align}
Applying the inverse complementary error function to the inequality constraints supports the transformation of the bound-based constraints into distance constraints as $d_{k}\pc{\boldsymbol{\theta},P}\geq \sigma_w \pc{\text{erfc}^{-1}\pc{\rho_k}}, \ \text{for} \ k\in \mathcal{K}$.
With this, problem \eqref{opt:mmddt_problem1} can be written as in \cite{liu2021intelligent} by explicitly writing the distance expression in the constraint functions which yields 
\begin{align}
\label{opt:mmddt_ris}
    &\min_{\boldsymbol{\theta}, P\in \mathbb{R}_+}  \ P \\
    &\hspace{0.7em}\text{s.t.}\hspace{0.5em} \PM{\pr{\boldsymbol{\theta}}_n}^2=1,\ \text{for} \ n \in \mathcal{N}, \hspace{1em} r_k=\sqrt{P}  \boldsymbol{h}_k^H \boldsymbol{\theta} \text{e}^{-j \text{arg} (s_k)}, \notag \\
    &\hspace{2.7em} \text{Re}\{r_k\} \sin{\phi}-\PM{\text{Im}\{r_k\}} \cos{\phi}\geq \alpha_k, \ \text{for} \ k\in \mathcal{K},\notag 
\end{align}
where $\alpha_k=\sigma_w\pc{ \text{erfc}^{-1}\pc{\rho_k}}$.
Since the MDDT constraint functions are upper bounds on the UBSEP constraints, one can understand the problem from \cite{liu2021intelligent} as a restricted version of the proposed PHUBSEP formulation. This implies that, for attaining the same SEP requisite the optimal transmit power minimization under MDDT constraints is greater or equal to the transmit power of the proposed approach.
\subsection{\textcolor{r3}{Convexity of the High-Resolution Constraint Functions}}
\label{app:high_res}
\textcolor{r2}{This section examines the SEP and UBSEP constraint functions formulated for high-resolution cases, proves that the SEP functions are convex, and demonstrates the conditions for convexity UBSEP functions. }

First consider the SEP constraint function $f_k(\boldsymbol{\Theta})=-\sum_{\xi=1}^2 \text{ln}\pc{\Phi\pc{\sqrt{\frac{P}{\sigma_w^2}}\ \text{tr}\pc{\boldsymbol{\Theta}  \boldsymbol{H}_{\xi,k} }} }-\beta_k$. \textcolor{r2}{As proven in Appendix \ref{app:sep}, the function $g(x)=-\ln\pc{\Phi\pc{\sqrt{\frac{P}{\sigma_w^2}} \ x}}-\beta_k$ is convex.} Note that, by definition $\text{tr}\pc{\boldsymbol{\Theta}  \boldsymbol{H}_{\xi,k} }=\sum_{i=1}^2 \sum_{j=1}^N {\theta}_{i,j} {h}_{i,j}^\xi $, where $\theta_{i,j}$ and ${h}_{i,j}^\xi$ denote the entry on the $i$-th row and $j$-th column of $\boldsymbol{\Theta}$ and $\boldsymbol{H}_{\xi,k}$, respectively. With this, $f_k(\boldsymbol{\Theta})=g(\sum_{i=1}^2 \sum_{j=1}^N {\theta}_{i,j} {h}_{i,j}^\xi )-\beta_k$ is a composition of the convex nondecreasing function $g$ with a linear function $\text{tr}\pc{\boldsymbol{\Theta}  \boldsymbol{H}_{\xi,k} }$, which yields a convex function \cite[Section 3.2.4]{Boyd_2004}. 

\textcolor{r2}{A similar path can be taken to derive the conditions of convexity of the UBSEP functions $f_k(\boldsymbol{\Theta})=\sum_{\xi=1}^2 \frac{1}{2}\text{erfc}\pc{ {\sqrt{\frac{P}{\sigma_w^2}}}\ \text{tr}\pc{\boldsymbol{\Theta}  \boldsymbol{U}_{\xi,k} } } -\rho_k$.} Note, however, that function $g(x)=\frac{1}{2}\text{erfc}\pc{ {\sqrt{\frac{P}{\sigma_w^2}}}x}-\rho_k$ is convex for only $x\geq0$. With this, the composed function $f_k(\boldsymbol{\Theta})=g(\sum_{i=1}^2 \sum_{j=1}^N {\theta}_{i,j} {u}_{i,j}^\xi )-\rho_k$, with ${u}_{i,j}^\xi$ denoting the entry on the $i$-th row and $j$-th column of $\boldsymbol{U}_{\xi,k}$ is convex in the regions where $\sum_{i=1}^2 \sum_{j=1}^N {\theta}_{i,j} {u}_{i,j}^\xi \geq 0$. \textcolor{r2}{This implies that $f_k(\boldsymbol{\Theta})$ is convex for $\text{tr}\pc{\boldsymbol{\Theta} \boldsymbol{U}_{\xi,k}}\geq0,\  \text{for}\ k \in \mathcal{K}, \ \xi \in\{1,2\}$.}

\ifCLASSOPTIONcaptionsoff
  \newpage
\fi

\bibliographystyle{IEEEtran}
\bibliography{bib-refs}

\begin{IEEEbiography}[{\includegraphics[width=1in,height=1.25in,clip,keepaspectratio]{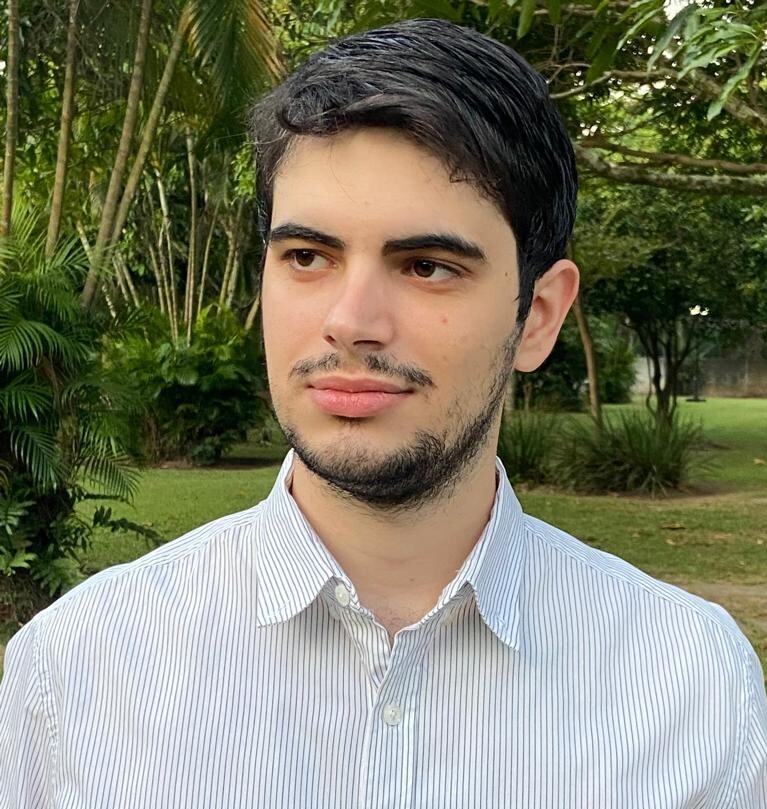}}]{Erico S. P. Lopes}
(S'21) received the degree in Electrical Engineering with emphases on Telecommunications and Electronics from the Pontifical Catholic University of Rio de Janeiro (PUC-Rio) (2019), a master's degree in Electrical Engineering from the same institution in the field of Telecommunications Systems (2021), and a Ph.D. in Electrical Engineering, also from PUC-Rio (2025). Since 2024, he has been with the National Institute of Industrial Property (INPI), where he serves as an industrial property researcher, responsible for patent examination in the field of telecommunications. His main areas of expertise include signal processing, wireless communications, and optimization. His research interests encompass massive MIMO systems, MIMO systems with low-resolution quantizers, beamforming, intelligent reflecting surfaces, and convex and non-convex optimization.
\end{IEEEbiography}

\begin{IEEEbiography}[{\includegraphics[width=1in,height=1.25in,clip,keepaspectratio]{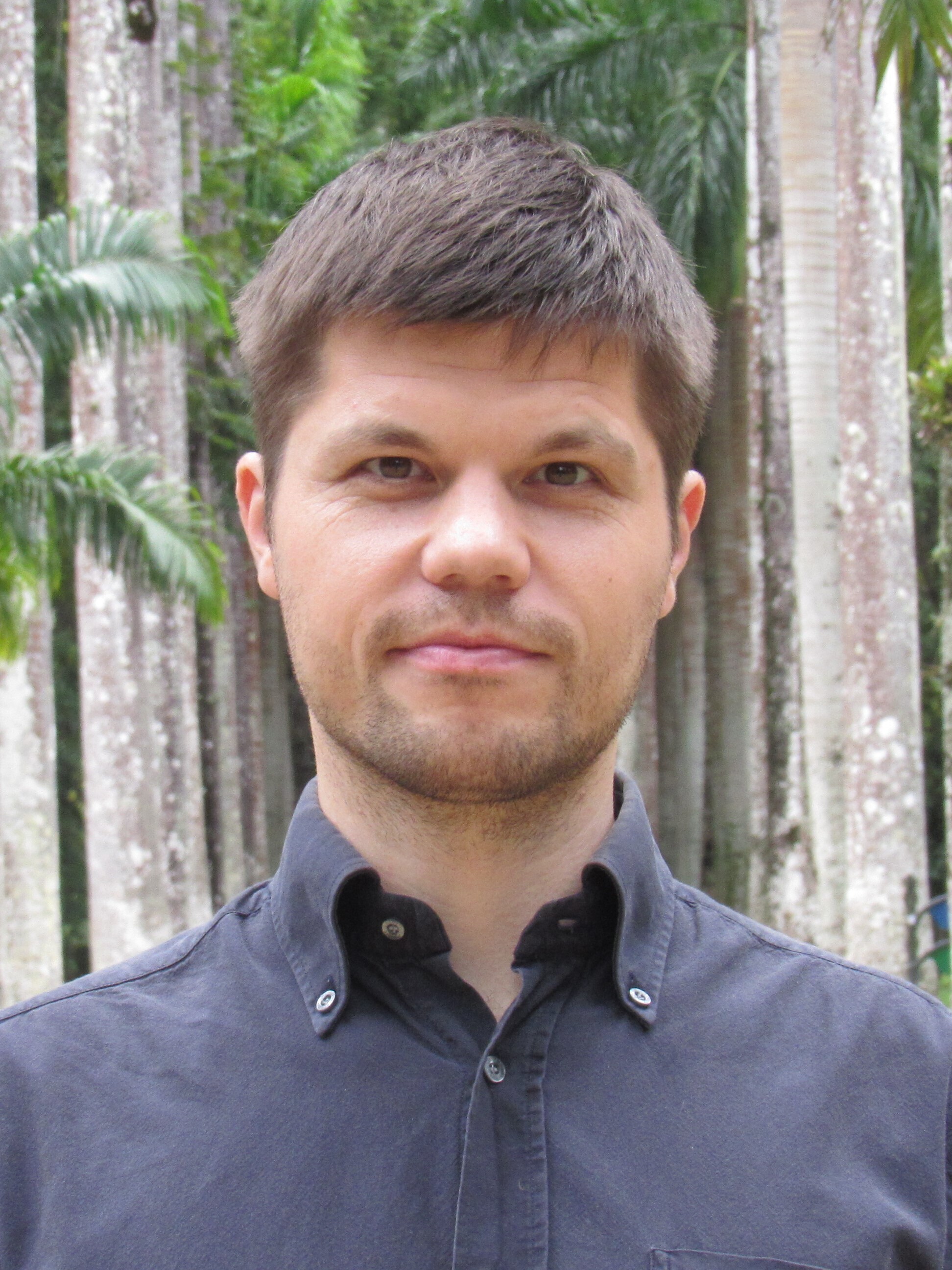}}]{Lukas T. N. Landau}
(S’13-M’17-SM'22) received the B.Sc.\ and the M.Sc.\ degrees in electrical engineering and information technology from the Ilmenau University of Technology, Germany, in 2009 and 2011, respectively, and the Ph.D.\ degree in electrical engineering and information technology from Technische Universit\"{a}t Dresden, Germany, in 2016. Since 2016, he has been with the Pontifical
Catholic University of Rio de Janeiro, Brazil, where he is currently an Adjunct Professor.
He also serves as an Associate Editor for IEEE Transactions on Wireless Communications, IEEE Wireless Communications Letters, the {\it{EURASIP Journal on Wireless Communications and Networking}} and {\it{Wireless Personal Communications}}.
His research interests lie in communications and signal processing.
\end{IEEEbiography}

\begin{IEEEbiography}[{\includegraphics[width=1in,height=1.25in,clip,keepaspectratio]{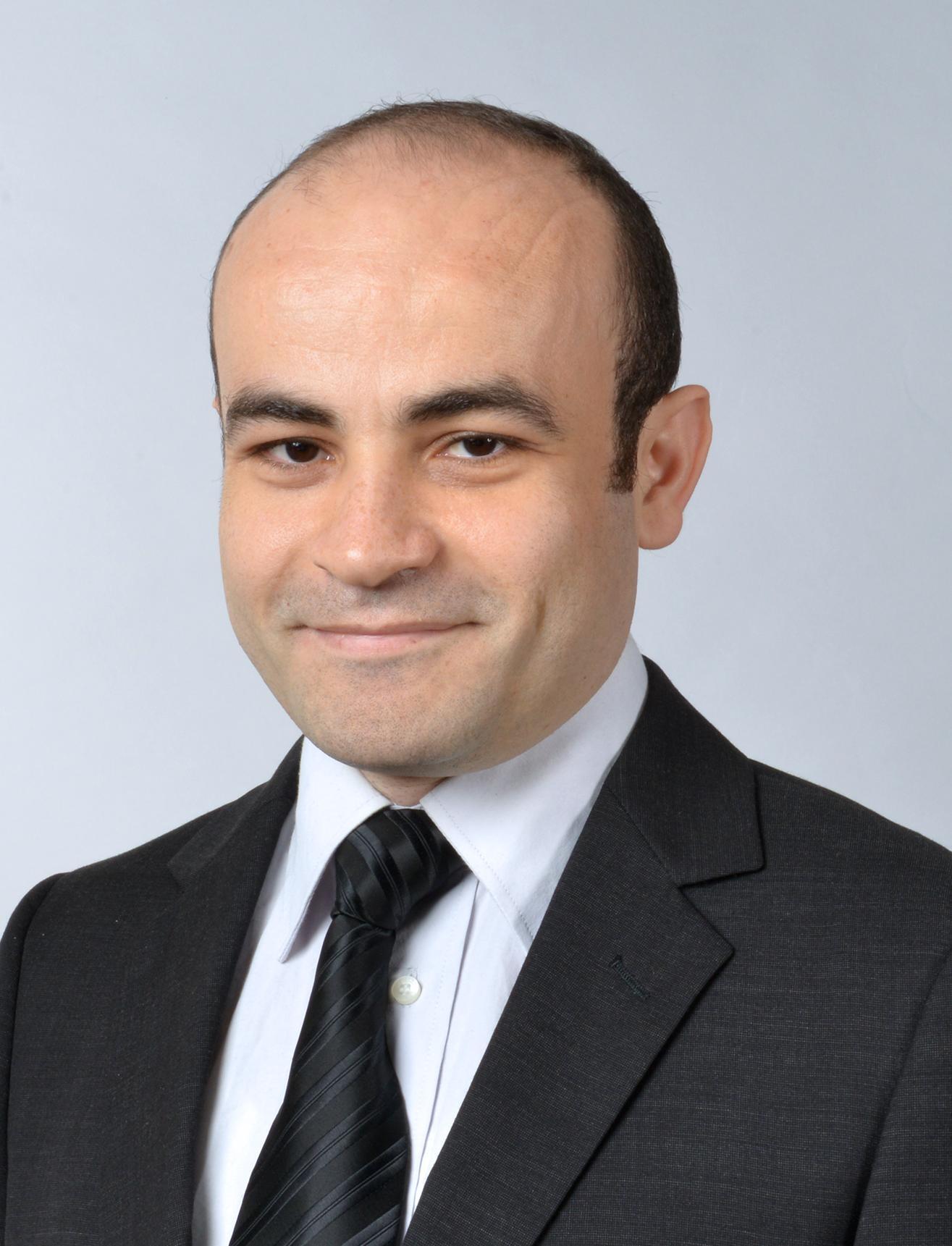}}]{Amine Mezghani}
(Member, IEEE) received the Ph.D. degree in electrical engineering from the Technical University of Munich, Germany, in 2015. Currently, he is an Assistant Professor with the Department of Electrical and Computer Engineering, University of Manitoba, Canada. He was a Postdoctoral Fellow with the University of Texas at Austin, USA, and a Postdoctoral Scholar with the Department of Electrical Engineering and Computer Science, University of California, Irvine, CA, USA. His research interests include millimeter-wave communications, massive MIMO, hardware constrained communication systems, antenna theory, and large-scale signal processing algorithms. He was the recipient of 2021 IEEE Signal Processing Society Best Paper Award, the 2023 Winnipeg Rh Institute Foundation Award for outstanding research accomplishments, and the 2016 joint Rohde \& Schwarz and EE department Outstanding Dissertation Award. He has published more than a hundred papers, particularly on the topic of signal processing and communications with low-resolution analog-to-digital and digital-to-analog converters.
\end{IEEEbiography}

\end{document}